\documentclass[12pt,a4paper]{article}
\usepackage{jheppub}
\usepackage{dsfont}
\usepackage{amssymb,amsmath}
\usepackage{epsfig}
\usepackage{epstopdf}
\usepackage{latexsym}
\usepackage{graphicx}
\usepackage{subfigure}
\usepackage{booktabs}
\usepackage{bbm}
\pdfoutput=1
\makeatletter
\def\@fpheader{\relax}
\makeatother

\def\O{{\cal O} }
\def\C{{\cal C} }
\def\A{{\cal A} }
\def\D{{\cal D} }
\def\W{{\cal W} }
\def\N{{\cal N} }
\def\S{{\cal S} }
\def\L{{\cal L} }
\def\P{{\cal P} }
\def\E{{\cal E} }
\def\H{{\cal H} }
\def\tr{{\text{tr}} }
\def\dcut{{%
    \setbox0\hbox{D}%
    \rlap{\hbox to \wd0{\hss ~/ \hss}}\box0}}

\subheader{\begin{flushright}
UTTG-23-12\\
TCC-022-12
\end{flushright}}

\title{Strongly Coupled Gauge Theories: High and Low Temperature Behavior of Non-local Observables}
\author{Willy Fischler,~ Sandipan Kundu}
\affiliation{Theory Group, Department of Physics, University of Texas, Austin, TX 78712, USA}
\affiliation{Texas Cosmology Center, University of Texas, Austin, TX 78712, USA}
\emailAdd{fischler@physics.utexas.edu}
\emailAdd{sandyk@physics.utexas.edu}

\abstract{We explore the high and low temperature behavior of non-local observables in strongly coupled gauge theories that are dual to AdS. We develop a systematic expansion for equal time two-point correlation, spatial Wilson loops and entanglement entropy at finite temperature using the AdS/CFT correspondence, leading to analytic expressions for these observables at high and low temperature limits. This approach enables the identification of the contributions of different regions of the bulk geometry to these gauge theory observables.}

\arxivnumber{1212.2643}

\begin{document}

\maketitle
\flushbottom

\section{Introduction}
The finite temperature behavior of non-abelian gauge theories has been of interests to physicists over the years (see \cite{Dine:1980jm, Gross:1980br, Weldon:1982aq, Landsman:1986uw, Braaten:1989mz, Witten:1998zw, Rey:1998bq, Shuryak:2008eq, Panero:2009tv, Megias:2010ku, Mojaza:2010cm, Petreczky:2012rq} and references therein). Of particular interest is the behavior at strong coupling where we have a powerful tool in the AdS/CFT correspondence \cite{Maldacena:1997re, Witten:1998qj, Gubser:1998bc, Aharony:1999ti}. This correspondence has been used in various areas ranging from Relativistic Heavy Ion Collider (RHIC) to condensed matter physics. The study of non-local observables (e.g. two-point function, Wilson loop, entanglement entropy) at finite temperature for strongly coupled gauge theories using the AdS/CFT correspondence is an interesting problem in its own right. One of the questions we focused on, is the regions in the bulk that contribute most significantly to these observables of the `boundary' gauge theory in the different limits. This led us to construct analytic expansions both at high and low temperature. 

In this paper, we will mainly consider strongly coupled large-N gauge theories in $d$-dimensions that are dual to AdS$_{d+1}$. As mentioned above, our goal is to understand analytically the behavior of non-local observables like equal time two-point function, spatial Wilson loop and entanglement entropy in the low and high temperature limits. We will henceforth develop a systematic expansion using the AdS/CFT dictionary. All of these observables of the boundary gauge theory, associated to a region A can be computed by calculating the area of the bulk extremal surface anchored on the boundary of A (for the two-point function the relevant  quantity is the geodesic, for what follows we will use the label `extremal surface' for the sake of simplicity). As a consequence of conformal invariance the dimensionless parameter of interest is $T l$, when we are computing some observable with length scale $l$ at temperature $T$. 

We develop separate expansion techniques at low and high temperature. At low temperature, i.e. $T\ll 1/l$, the extremal surface is restricted to be near the boundary region and thus the leading contribution to the area comes from the AdS-boundary. This contribution is just the zero temperature result which has been thoroughly investigated and well understood in the literature. Finite temperature corrections correspond to the deviation of the bulk geometry from pure AdS. At low temperature, these corrections are small and can be computed perturbatively. On the other hand, at high temperature, i.e. $T\gg 1/l$, the physics is completely different and more interesting. As the thermal fluctuations become more and more significant, the extremal surface associated with the observable approaches the horizon.\footnote{As recently discussed in \cite{Hubeny:2012ry}, extremal surfaces of any dimensionality and anchored on arbitrary shaped region on the boundary at constant time can not penetrate the horizon in any static spherically symmetric space times. At high temperature, we will show that the extremal surface approaches the horizon exponentially fast but it always stays finite distance above the horizon.} At high temperature, the extremal surface tends to wrap a part of the horizon and the leading contribution comes from the near horizon region of the surface. This has following consequences: 1) the equal time two-point function decays exponentially; 2) the leading contribution to the entanglement entropy (spatial Wilson loop) is proportional to the volume (area) of the associated region. The sub-leading terms are more complicated because they receive contributions from full bulk geometry. The sub-leading terms in various observables, in particular the entanglement entropy, may possibly contain important information about these gauge theories.

Fascinating question arises when one considers the high temperature regime of the field theory. At weak coupling, it is known that the high temperature behavior of the gauge invariant correlation functions  (e.g. $\langle \tr F_{\mu \nu}^2 (l) \tr F_{\mu \nu}^2 (0)\rangle$, spatial Wilson loop) is given by classical statistical mechanics\cite{Dine:1980jm}, this obviously does not apply to thermodynamic quantities like the specific heat. Indeed, one expects on the gauge theory side, that the leading contributions to the gauge invariant correlation functions with dependence on one length scale $ l$ come from modes with wavelength $\sim l$. As the system is heated the modes with wavelength $\sim l$ become more and more populated. Therefore, equal time two-point functions and the spatial Wilson loops should behave classically at high temperature ($T\gg 1/l$). It is tempting to speculate that at strong coupling an analogous statement can be made. This raises a puzzling question. On the one hand, the high temperature behavior is given by the near horizon geometry and on the other hand we have speculated that it is given by classical statistical mechanics. Can we then conclude that the near horizon contribution corresponds to classical statistical mechanics on the boundary?
 
In contrast, we found that it is easier to understand the high temperature behavior of entanglement entropy, which is a fully quantum mechanical concept and does not have a classical analog. At high temperature ($T\gg 1/l$), the leading finite contribution to the entanglement entropy comes from the near horizon region and it is just the thermal entropy. Whereas, contributions arising from deviations away from the near horizon region are responsible for the sub-leading terms, which actually measure quantum entanglement between the region and its surroundings. This connection between the near horizon region of the bulk and no quantum entanglement in the boundary theory is possibly a manifestation of UV/IR duality\cite{Susskind:1998dq}.

The rest of the paper is organized as follows. We start with a brief review of how the temperature of the boundary gauge theory is related to the horizon of the bulk theory in section \ref{sectemp}. In section \ref{sec2pt}, we develop a systematic expansion for the equal time two-point function at finite temperature. Then using that expansion we examine the high and low temperature behavior of the two-point function.  In section \ref{secwl} and \ref{secee}, we go on to develop similar systematic expansions for rectangular spatial Wilson loop and entanglement entropy of a rectangular strip respectively at finite temperature. Consequently, we obtain analytic expressions for the spatial Wilson loop and  the entanglement entropy at high and low temperature limits. In section \ref{hvb} we will generalize these techniques for non-relativistic theories with hyperscaling violation and show that the physics is exactly the same. Finally, we summarize all the results in section \ref{secex} by considering the prototype case of four-dimensional $\N=4$ Super Yang-Mills gauge theory. Then we conclude in section \ref{conclusions} with future directions. Several technical details have been relegated to two appendices.

%%%%%%%%%%%%%%%%%%%%%%%%%%%%%%%%%%%%%%%%%%%%%%%%%%%%%%%%%%%%%%%%%%%%%%%%%%%%%%%%%%
\section{Gauge theories at finite temperature: Schwarzschild-AdS$_{d+1}$}\label{sectemp}
In this paper, we will mainly consider strongly coupled large-N gauge theories in $d-$dimensions that are dual to AdS$_{d+1}$. At nonzero temperature, all the thermal effects can be analyzed simply by introducing a black hole in the bulk. In other words, at finite temperature AdS$_{d+1}$ should be replaced by Schwarzschild-AdS$_{d+1}$  
\begin{equation}\label{sads}
ds^2=-\frac{r^2}{R^2}\left(1-\frac{r_H^d}{r^d}\right)dt^2+\frac{r^2}{R^2} d\vec{x}^2+\frac{R^2}{r^2\left(1-\frac{r_H^d}{r^d}\right)}dr^2, 
\end{equation}
where, $R$ is the AdS radius. The temperature $T$ is given by the Hawking temperature of the black hole which can be determined by demanding that the Euclidean continuation of the metric (\ref{sads})
\begin{equation}
ds^2=\frac{r^2}{R^2}\left(1-\frac{r_H^d}{r^d}\right)dt_E^2+\frac{r^2}{R^2} d\vec{x}^2+\frac{R^2}{r^2\left(1-\frac{r_H^d}{r^d}\right)}dr^2, 
\end{equation}
is regular at the horizon. Near the horizon $r=r_H$, the Euclidean metric looks like
\begin{equation}
ds_{\text{near horizon}}^2=\rho^2 d\phi^2+d\rho^2+\frac{r_H^2}{R^2} d\vec{x}^2, 
\end{equation}
where $\rho$ and $\phi$ are defined as
\begin{equation}\label{nearhorizon}
\rho=2\sqrt{\frac{R^2(r-r_H)}{r_H d}}, \qquad \phi=\frac{r_H d}{2 R^2}t_E.
\end{equation}
Now $\phi$ must be periodic with period $2\pi$ in order to avoid a conical singularity at $\rho=0$ (i.e. $r=r_H$). From equation (\ref{nearhorizon}) it is clear that Euclidean time $t_E$ is also periodic and the period can be identified with $1/T$. Therefore, the temperature of the $d$-dimensional boundary field theory (after restoring $\hbar$) is given by,
\begin{equation}\label{temperature}
T=\frac{\hbar ~r_H d}{4 \pi R^2}.
\end{equation}

%%%%%%%%%%%%%%%%%%%%%%%%%%%%%%%%%%%%%%%%%%%%%%%%%%%%%%%%%%%%%%%%%%%%%%%%%%%%%%%
\section{Two-point function}\label{sec2pt}
We will compute the equal time two-point function of some gauge-invariant scalar operator $\O(t,x)$ in the limit $N\gg 1, \lambda\equiv g_{YM}^2 N\gg 1$. According to the AdS/CFT correspondence, in this limit, the calculation of two-point function \cite{Gubser:1998bc, Witten:1998qj} of gauge-invariant operator $\O(t,x)$ simply reduces to solving classical equation of motion for some bulk field that couples to $\O(t,x)$. This prescription has been successfully generalized \cite{Son:2002sd, Herzog:2002pc} for the finite temperature case (for a nice review see \cite{Son:2007vk}). 

For our purpose it is more convenient to use the alternate prescription proposed in \cite{Banks:1998dd}. Following  \cite{Balasubramanian:1999zv} the equal-time two-point function $\langle \O(t,x)\O(t,y)\rangle$ can be represented as a path integral that sums over all the paths that join the boundary points $(t,x)$ and $(t,y)$  
\begin{equation}
\langle \O(t,x)\O(t,y)\rangle=\int \D \P e^{-\Delta L(\P)},
\end{equation}
where $L(\P)$ is the proper length of the path and $\Delta$ is the conformal dimension of the operator $\O$. For operators with large conformal dimension, we can perform a saddle point approximation\footnote{It is important to note that for Lorentzian correlators this approximation should be used more carefully \cite{Louko:2000tp,Fidkowski:2003nf, Festuccia:2005pi}.}
\begin{equation}
\langle \O(t,x)\O(t,y)\rangle\approx \sum_{\text{geodesics}} e^{-\Delta \L},
\end{equation}
where $\L$ is the geodesic length of the geodesic between boundary points $(t,x)$ and $(t,y)$. The geodesic length $\L$ has a divergence that comes from the boundary. This divergence can be removed by introducing a cutoff $r_b$ and then defining renormalized geodesic length by
\begin{equation}\label{rengeo}
\L_{ren}=\L-2\ln r_b.
\end{equation}
Note that the divergent piece is independent of both temperature and dimension. Finally, the renormalized two point function is given by,
\begin{equation}\label{twopoint}
\langle \O(t,x)\O(t,y)\rangle\approx  e^{-\Delta \L_{ren}}.
\end{equation}

\subsection{Finite temperature expansion}
The bulk-metric is given by\footnote{For the sake of cleanliness, we will use AdS radius $R=1$. We will restore $R$ by dimensional analysis whenever necessary.}
\begin{equation}\label{metric}
ds^2=-r^2\left(1-\frac{r_H^d}{r^d}\right)dt^2+r^2 d\vec{x}^2+\frac{1}{r^2\left(1-\frac{r_H^d}{r^d}\right)}dr^2. 
\end{equation}
We can choose our coordinates so that the two points are $(t,x=-\frac{l}{2},0,...)$ and  $(t,x=\frac{l}{2},0,...)$ respectively. Therefore, the relevant part of the Schwarzschild-AdS$_{d+1}$ metric is:
\begin{equation}
ds^2=r^2 dx^2+\frac{1}{r^2\left(1-\frac{r_H^d}{r^d}\right)}dr^2. 
\end{equation}
In the affine parametrization, the geodesic equations are given  by,
\begin{align}
\dot{x}=& \frac{r_c}{r^2},\label{xeqn}\\
\dot{r}=& \pm r \sqrt{\left(1- \frac{r_c^2}{r^2}\right)\left(1-\frac{r_H^d}{r^d}\right)},\label{reqn}
\end{align}
where, $r_c$ is an integral of motion associated with the Killing vector $\partial_x$ and $r=r_c$ represents the point of closest approach of the geodesic. We also have taken the affine parameter to represent the geodesic proper length $s$. It is clear from equation (\ref{reqn}) that each geodesic has two branches, which we will denote as $x_+(r)$ and $x_-(r)$, joined smoothly at $(r=r_c, x=0)$. In principle, $r_c$ can be determined using the boundary conditions:
\begin{equation}
x_-(\infty)=-\frac{l}{2}, \qquad  x_+(\infty)=\frac{l}{2}.
\end{equation}
Now using equations (\ref{xeqn}, \ref{reqn}), for the positive branch of the geodesic, we obtain
\begin{equation}
\frac{dr}{dx}= \frac{r^3}{r_c} \sqrt{\left(1- \frac{r_c^2}{r^2}\right)\left(1-\frac{r_H^d}{r^d}\right)}.\label{rxeqn}
\end{equation}
Integration of the last equation leads to
\begin{align}
\frac{l}{2}=&\int_{r_c}^{\infty}\frac{r_c dr}{r^3 \sqrt{\left(1- \frac{r_c^2}{r^2}\right)}} \left(1-\frac{r_H^d}{r^d}\right)^{-1/2}\nonumber\\
=& \frac{1}{r_c}\int_{0}^{1}\frac{u du}{ \sqrt{1- u^2}} \left(1-\frac{r_H^d}{r_c^d}u^d\right)^{-1/2}. \label{rceqn}
\end{align}
Unfortunately this integration can be performed analytically only for Schwarzschild-AdS$_3$ (i.e. $d=2$). For a general $d$, we can do a systematic expansion.
\begin{align}
\frac{l}{2}=& \frac{1}{r_c}\int_{0}^{1}\frac{u du}{ \sqrt{1- u^2}}\sum_{n=0}^\infty\frac{\Gamma\left[\frac{1}{2}+n\right]}{\sqrt{\pi } \Gamma[1+n]}\left(\frac{r_H}{r_c}\right)^{nd}u^{nd}\\
=&\frac{1}{2 r_c}\sum_{n=0}^\infty\frac{\Gamma\left[\frac{1}{2}+n\right]\Gamma\left[1+\frac{nd}{2}\right]}{ \Gamma[1+n]\Gamma\left[\frac{3}{2}+\frac{nd}{2}\right]}\left(\frac{r_H}{r_c}\right)^{nd}.\label{seriesrc}
\end{align}
We should be more careful about the convergence of the series (\ref{seriesrc}) before we use it. For large $n$, the series goes as $\sim \frac{1}{n}\left(r_H/r_c\right)^{nd}$ and hence the series converges for $r_H/r_c<1$. For any finite temperature, it can be shown \cite{Hubeny:2012ry} that $r_c>r_H$. Therefore, the sum (\ref{seriesrc}) is well-defined.

Next we will calculate the regularized geodesic length by using equation (\ref{reqn})
\begin{align}
\L=&2 \int_{r_c}^{\infty}\frac{ dr}{r \sqrt{\left(1- \frac{r_c^2}{r^2}\right)}} \left(1-\frac{r_H^d}{r^d}\right)^{-1/2}. \label{seqn}
\end{align}      
Where, the factor of 2 comes because of the two branches of the geodesic. The equation (\ref{seqn}) has a divergence that comes from the upper limit of the integration. This divergence can be removed by introducing a cutoff $r_b$ and then using equation (\ref{rengeo}) to define the renormalized geodesic length. Now we will proceed to develop a systematic expansion for $\L_{ren}$
\begin{align}
\L_{ren}=& 2\int_{r_c/r_b}^{1}\frac{du}{u \sqrt{1- u^2}} \left(1-\frac{r_H^d}{r_c^d}u^d\right)^{-1/2}-2\ln r_b\\
=& 2 \int_{r_c/r_b}^{1}\frac{du}{u \sqrt{1- u^2}}\sum_{n=0}^\infty\frac{\Gamma\left[\frac{1}{2}+n\right]}{\sqrt{\pi } \Gamma[1+n]}\left(\frac{r_H}{r_c}\right)^{nd}u^{nd}-2\ln r_b\\
=& 2 \int_{r_c/r_b}^{1}\frac{du}{u \sqrt{1- u^2}}-2\ln r_b+2 \int_{0}^{1}\frac{du}{u \sqrt{1- u^2}}\sum_{n=1}^\infty\frac{\Gamma\left[\frac{1}{2}+n\right]}{\sqrt{\pi } \Gamma[1+n]}\left(\frac{r_H}{r_c}\right)^{nd}u^{nd},
\end{align}
where, we have used the fact that only $n=0$ term is divergent. Finally, we have
\begin{align}
\L_{ren}=& 2\ln\left(\frac{2}{r_c}\right)+\sum_{n=1}^\infty\frac{\Gamma\left[\frac{1}{2}+n\right]\Gamma\left[\frac{nd}{2}\right]}{ \Gamma[1+n]\Gamma\left[\frac{1}{2}+\frac{nd}{2}\right]}\left(\frac{r_H}{r_c}\right)^{nd}.\label{sreneqn}
\end{align}
Again it can be shown that for large $n$, the infinite series in equation (\ref{sreneqn}) goes as $\sim \frac{1}{n}\left(r_H/r_c\right)^{nd}$ and hence the series converges for $r_H/r_c<1$. In principle, the rest of the procedure is very simple; we have to solve equation (\ref{seriesrc}) for $r_c$ and then use that $r_c$ in equation (\ref{sreneqn}) to get the renormalized geodesic length.\footnote{We would like to stress that the equations (\ref{seriesrc}, \ref{sreneqn}) are valid for any temperature.} The renormalized two-point function is then given by equation (\ref{twopoint}). But in practice, this procedure can be performed exactly only for d=2. For $d\neq 2$, it is not possible to solve equation (\ref{seriesrc})  analytically to find $r_c$ as a function of $l$. However at low temperature (i.e. $r_H l \ll 1$), we can compute $\L_{ren}$ perturbatively using equations (\ref{seriesrc}, \ref{sreneqn}). And more interestingly, equation (\ref{sreneqn}) can also be used to determine the high temperature (i.e. $r_H l\gg 1$) behavior of the two-point function.

\subsection{Two-point function: CFT in $(1+1)$ dimensions}
A trivial example, as mentioned earlier, is Schwarzschild-AdS$_{3}$ (i.e. $d=2$). For $d=2$, equation (\ref{seriesrc}) reduces to
\begin{align}
\frac{l}{2}=\frac{1}{ r_c}\sum_{n=0}^\infty\frac{1}{2n+1}\left(\frac{r_H}{r_c}\right)^{2n}=\frac{\tanh ^{-1}\left(\frac{r_H}{r_c}\right)}{r_H}.
\end{align}
Therefore,
\begin{equation}
r_c=\frac{r_H}{\tanh\left(\frac{ r_H l}{2}\right)}.
\end{equation}
And equation (\ref{sreneqn}) becomes
\begin{align}
\L_{ren}=& 2\ln\left(\frac{2}{r_c}\right)+\sum_{n=1}^\infty\frac{1}{n}\left(\frac{r_H}{r_c}\right)^{2n}\nonumber\\
=&2\ln\left(\frac{2}{r_c}\right)-\ln \left(1-\frac{r_H^2}{r_c^2}\right)\nonumber\\
=& -\ln \left[\frac{r_H^2}{2\left(\cosh(r_H l)-1\right)}\right].
\end{align}
Therefore,
\begin{equation}
\langle \O(t,x)\O(t,y)\rangle=r_H^{2\Delta} \left[\frac{1}{2\left(\cosh(r_H l)-1\right)}\right]^\Delta
\end{equation}
where, $l=|x-y|$. Now using equation (\ref{temperature}) we finally have
\begin{equation}
\langle \O(t,x)\O(t,y)\rangle=(\pi T)^{2\Delta} \left[\frac{2}{\cosh\left(2 \pi T |x-y|\right)-1}\right]^\Delta.
\end{equation}
%%%%%%%%%%%%%%%%%%%%%%%%%%%%%%%%%%%%%%%%%%%%%%%%%%%%%%%%%%%%%%%%%%%%%%%%%%%%%%%%%%%%%%%%%

\subsection{Low temperature two-point function}
At low temperature, $r_c\gg r_H$ and hence the leading contribution to the geodesic length comes from the boundary. The boundary is still AdS and we should get the zero temperature two-point function as the leading term. Finite temperature corrections can be computed by considering deviations from the boundary geometry. At low temperature ($T l\ll 1$), the corrections to zero temperature result are small and hence can be computed perturbatively. From equation (\ref{seriesrc}), keeping only few subleading terms, we get

\begin{align}
l=\frac{1}{r_c}\left[2+ \frac{\sqrt{\pi } \Gamma \left(\frac{d}{2}+1\right)}{2 \Gamma \left(\frac{d+3}{2}\right)}\left(\frac{r_H}{r_c}\right)^d+\frac{3 \sqrt{\pi } \Gamma (d+1)}{8 \Gamma \left(d+\frac{3}{2}\right)}\left(\frac{r_H}{r_c}\right)^{2d}+...\right].
\end{align} 
Solving the last equation perturbatively and then using that solution in equation (\ref{sreneqn}), we get a perturbative expression for $\L_{ren}$ (details of the calculation are relegated to appendix \ref{lte}). Finally using (\ref{temperature}), for the $d$-dimensional boundary theory at low temperature we obtain
\begin{align}
\langle \O(t,x)\O(t,y)\rangle=\frac{1}{|x-y|^{2 \Delta}}&\left[1+\C_1 \left(\frac{2\pi T |x-y|}{d}\right)^d \right.\nonumber\\
&\left.+ \C_2 \left(\frac{2\pi T |x-y|}{d}\right)^{2d}+\O\left(\frac{2 \pi T |x-y|}{d}\right)^{3d}\right],
\end{align}
where
\begin{align}
\C_1=&-\frac{ \sqrt{\pi } \Delta  \Gamma \left(\frac{d}{2}\right)}{4 \Gamma \left(\frac{d+3}{2}\right)},\\
\C_2=&\frac{1}{64} \left[\pi  \Delta  \left(d^2+2 \Delta \right)\left(\frac{ \Gamma \left(\frac{d}{2}\right)}{\Gamma \left(\frac{d+3}{2}\right)}\right)^2-\frac{12 \sqrt{\pi } \Delta  \Gamma (d)}{\Gamma \left(d+\frac{3}{2}\right)}\right].
\end{align}
Note that $\C_1$ is negative as expected indicating a decrease in two-point correlation because of thermal fluctuations. 

Particularly for $d=4$, using equations (\ref{lowtemp2}-\ref{c2}), at low temperature limit $(T|x-y|\ll 1)$ we have
\begin{align}\label{ads4low}
\langle \O(t,x)\O(t,y)\rangle=\frac{1}{|x-y|^{2 \Delta}}&\left[1-\frac{2 \Delta }{15} \left(\frac{\pi T |x-y|}{2}\right)^4 \right.\nonumber\\
&\left.+ \frac{2 \Delta  (7 \Delta +26)}{1575} \left(\frac{\pi T |x-y|}{2}\right)^{8}+\O\left(\frac{\pi T |x-y|}{2}\right)^{12}\right].
\end{align}
%%%%%%%%%%%%%%%%%%%%%%%%%%%%%%%%%%%%%%%%%%%%%%%%%%%%%%%%%%%%%%%%%%%%%%%%%%%%%%%%%%%%%%%%%%
\subsection{High temperature two-point function}
At high temperature (i.e. $T l \gg 1$), it is more difficult to do a systematic expansion for the two-point function. Nevertheless, we can find out the asymptotic behavior of the two-point function easily. At high temperature, the leading contribution comes from the near horizon part of the geodesic and this piece is easy to compute. On the other hand, the full bulk contributes to the subleading terms and hence they are more complicated.

%We will compute the leading term by two different methods. First, we will compute the two-point function by approximating the geodesic. And then to get some confidence in our result, we will use the finite temperature expansion to reproduce the leading contribution.

At very high temperature, $r_c$ approaches $r_H$. The obvious guess of taking the limit $r_c\rightarrow r_H$ in equation (\ref{sreneqn}) does not work because the infinite series in equation (\ref{sreneqn}) converges only when $r_c< r_H$. But we can rewrite equation (\ref{sreneqn}) in a way that allows us to take the limit  $r_c\rightarrow r_H$ without encountering any divergence.

\begin{align}
\L_{ren}=& 2\ln\left(\frac{2}{r_c}\right)+\sum_{n=1}^\infty\frac{\Gamma\left[\frac{1}{2}+n\right]\Gamma\left[\frac{nd}{2}\right]}{ \Gamma[1+n]\Gamma\left[\frac{1}{2}+\frac{nd}{2}\right]}\left(\frac{r_H}{r_c}\right)^{nd}\nonumber\\
=& 2\ln\left(\frac{2}{r_c}\right)+\sum_{n=1}^\infty\left(\frac{1+n d}{nd}\right)\frac{\Gamma\left[\frac{1}{2}+n\right]\Gamma\left[1+\frac{nd}{2}\right]}{ \Gamma[1+n]\Gamma\left[\frac{3}{2}+\frac{nd}{2}\right]}\left(\frac{r_H}{r_c}\right)^{nd}.
\end{align}  
Now, using equation (\ref{seriesrc}), we obtain
\begin{align}
\L_{ren}=&  2\ln\left(\frac{2}{r_c}\right)+\left(r_c l -2\right)+\sum_{n=1}^\infty\left(\frac{1}{nd}\right)\frac{\Gamma\left[\frac{1}{2}+n\right]\Gamma\left[1+\frac{nd}{2}\right]}{ \Gamma[1+n]\Gamma\left[\frac{3}{2}+\frac{nd}{2}\right]}\left(\frac{r_H}{r_c}\right)^{nd}.\label{modsren}
\end{align}  
The term $-2$ in the last equation comes from $n=0$ term of the series (\ref{seriesrc}). The infinite series in equation (\ref{modsren}) converges even for $r_c=r_H$ and hence the limit $r_c\rightarrow r_H$ exists. At high temperature, $r_c\sim r_H$ and now the leading behavior can be determined by taking the limit $r_c\rightarrow r_H$ in equation (\ref{modsren})
\begin{equation}
\L_{ren}\approx  2\ln\left(\frac{2}{r_H}\right)+\left(r_H l -2\right)+\sum_{n=1}^\infty\left(\frac{1}{nd}\right)\frac{\Gamma\left[\frac{1}{2}+n\right]\Gamma\left[1+\frac{nd}{2}\right]}{ \Gamma[1+n]\Gamma\left[\frac{3}{2}+\frac{nd}{2}\right]}.
\end{equation}
Therefore, the high temperature two-point function is approximately given by,
\begin{align}\label{hightemp}
\langle \O(t,x)\O(t,y)\rangle\approx \A_{d,\Delta}~ r_H^{2 \Delta}~ e^{-\Delta r_H l}.
\end{align}
Where, the prefactor $\A_{d,\Delta}$ is a constant that depends on the dimension $d$ and the conformal dimension $\Delta$ of the operator $\O$
\begin{equation}
\A_{d,\Delta}=\left\{\frac{1}{4}\exp\left[2-\sum_{n=1}^\infty\left(\frac{1}{nd}\right)\frac{\Gamma\left[\frac{1}{2}+n\right]\Gamma\left[1+\frac{nd}{2}\right]}{ \Gamma[1+n]\Gamma\left[\frac{3}{2}+\frac{nd}{2}\right]}\right]\right\}^{\Delta}.
\end{equation}
Finally, replacing $r_H$ by corresponding temperature, for the $d-$dimensional boundary theory we obtain
\begin{align}\label{highleading}
\langle \O(t,x)\O(t,y)\rangle\approx \A_{d,\Delta}~ \left(\frac{4 \pi T}{d}\right)^{2 \Delta}~ e^{- 4\pi\Delta T |x-y|/d}.
\end{align}

\begin{figure}[tb!]
\centering
\includegraphics[width=0.6\textwidth]{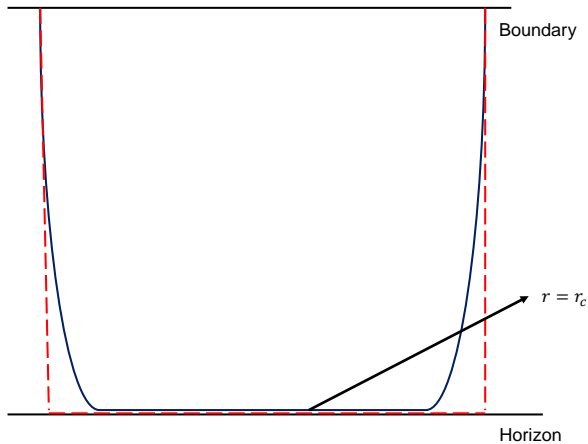}
\caption{At high temperature ($r_H l\gg 1$), the actual geodesic (solid blue line) can be approximated by the dashed red line curve that consists of $x=-l/2, r=r_H, x=l/2$.}
\label{approxgeodesic}
\end{figure}

The exponential decay of the two-point function at high temperature can be understood easily by looking at the geodesic for $r_H l\gg 1$. The actual U-shaped geodesic can be approximated by a curve that consists of $x=-l/2, r=r_H, x=l/2$ (see figure \ref{approxgeodesic}). As $r_H l$ is increased, the actual geodesic approaches the approximate one. But it can be shown that the proper length of this approximate curve is always greater than the actual geodesic and even in the limit $r_H l \rightarrow \infty$, two does not coincide. However in the high temperature limit, the most dominant contribution to the geodesic length comes from the near horizon part which can be reckoned from the approximate curve. The length of the near horizon part of the approximate curve is $S\sim r_H l$. Therefore, it is expected that the two-point function $\langle \O(t,x)\O(t,y)\rangle\sim e^{-\Delta r_H l}$.

We have computed the leading behavior of the high temperature ($Tl\gg 1$) two-point function by utilizing the fact that at high temperature, $r_c\sim r_H$. It is a good exercise to figure out exactly how close to the horizon the geodesic can reach. As shown in appendix \ref{subleading}, $r_c$ approaches $r_H$ exponentially fast
\begin{equation}\label{rc}
r_c=r_H\left(1+\E_d ~e^{-\sqrt{\frac{d}{2}}l r_H}+...\right),
\end{equation}
where $\E_d$ is a numerical constant given by,
\begin{equation}
\E_d=\frac{1}{d}\exp\left[\sqrt{\frac{d}{2}}\left\{2+\sum_{n=1}^\infty\left(\frac{\Gamma\left[\frac{1}{2}+n\right]\Gamma\left[1+\frac{nd}{2}\right]}{ \Gamma[1+n]\Gamma\left[\frac{3}{2}+\frac{nd}{2}\right]}-\frac{\sqrt{2}}{\sqrt{d}n}\right)\right\}\right].
\end{equation}
The geodesic always stays finite distance above the horizon which is consistent with \cite{Hubeny:2012ry}. Using equation (\ref{rc}), we can calculate the next order correction to the two-point function (for details see appendix \ref{subleading}). However, the subleading term is exponentially suppressed
\begin{align}\label{ht}
\langle \O(t,x)\O(t,y)\rangle = \A_{d,\Delta}~ \left(\frac{4 \pi T}{d}\right)^{2 \Delta}~ e^{- 4\pi\Delta T |x-y|/d}\left[1+\sqrt{\frac{2}{d}}\Delta \E_d ~e^{-2\sqrt{\frac{2}{d}}\pi T |x-y|}+...\right],
\end{align}
where the dots represent the higher order correction terms. Therefore, at high temperature, the two-point function is well approximated by equation (\ref{highleading}).

%%%%%%%%%%%%%%%%%%%%%%%%%%%%%%%%%%%%%%%%%%%%%%%%%%%%%%%%%%%%%%%%%%%%%%%%%%%%%%%
\section{Spatial Wilson loop}\label{secwl}
Wilson loops are another set of important gauge-invariant non-local observables in any gauge theory. In a gauge theory, the Wilson loop operator is a path ordered contour integral of the gauge field
\begin{equation}\label{wilsonloop}
W(\C)=\frac{1}{N}Tr\left(\P e^{\oint_{\C}A}\right),
\end{equation}
where the trace is over the fundamental representation and $\C$ denotes a closed loop in spacetime. Expectation values of the Wilson loops are useful to understand the non-perturbative behavior of non-Abelian gauge theories and have important applications to confinement/deconfinement transitions and quark screenings in QCD-like theories.

In the AdS/CFT correspondence, the expectation value of the Wilson loop\footnote{In these gauge theories the Wilson loop operators are typically given by some generalization of equation (\ref{wilsonloop}).} is related to the string partition function \cite{Maldacena:1998im}
\begin{equation}
\langle W(\C)\rangle=\int \D \Sigma ~e^{-S_{NG}(\Sigma)}, 
\end{equation}
where we integrate over all the string worldsheets $\Sigma$ with the boundary condition $\partial \Sigma=\C$ at the AdS boundary and $S_{NG}(\Sigma)$ corresponds to the Nambu-Goto action for the string worldsheet
\begin{equation}\label{nambugoto}
S_{NG}=\frac{1}{2 \pi \alpha '}\int d\tau d\sigma \sqrt{\det\left(g_{\mu \nu}\partial_\alpha x^{\mu}\partial_\beta x^{\nu}\right)}.
\end{equation}
Where $g_{\mu \nu}$ is the bulk metric and $1/2\pi \alpha'$ is the string tension; $x^{\mu}(\tau,\sigma)$ is the location of the string worldsheet in this $(d+1)$-dimensional spacetime. In the strong coupling limit, $\alpha'\ll 1$ and we can perform a saddle point approximation, yielding
\begin{equation}
\langle W(\C)\rangle= e^{-S_{NG}(\Sigma_0)}, 
\end{equation} 
where $\Sigma_0$ represents the minimal area surface with the boundary condition $\partial \Sigma_0=\C$.
%%%%%%%%%%%%%%%%%%%%%%%%%%%%%%%%%%%%%%%%%%%%%%%%%%%%%%%%%%%%%%%%%%%%%%%%%%%%%%%%%%%%%%%%%%%%%%%%%%%%%%%%%%%%%%%%%
\subsection{Rectangular Wilson loop: Finite temperature expansion}
In this section, we will consider spatial rectangular Wilson loops specified by $x\equiv x^1 \in \left[-\frac{l}{2},\frac{l}{2}\right],~ y\equiv x^2\in \left[-\frac{L}{2},\frac{L}{2}\right]$ and $t=x^i=0$ for $i > 2$ (see figure \ref{holo} for a pictorial representation). And we will assume that $L \gg l$. The two dimensional worldsheet can be parameterized by the coordinates $\sigma^\alpha \equiv (\tau, \sigma)$.  We will pick $\sigma=r,~\tau=y$ as the worldsheet coordinates. In the limit $L\rightarrow\infty$, the extremal surface is translationally invariant along the $y$-direction. Therefore, only coordinate $x=x(r)$ has a nontrivial profile and the induced metric on the worldsheet is given by
\begin{equation}
ds^2_{ws}=r^2 dy^2+ \left[r^2 x'^2+ \frac{1}{r^2\left(1-\frac{r_H^d}{r^d}\right)}\right]dr^2
\end{equation}
and the action (\ref{nambugoto}) becomes
\begin{equation}
S_{NG}=\frac{L}{2 \pi \alpha '}\int dr \sqrt{r^4 x'^2+ \frac{1}{\left(1-\frac{r_H^d}{r^d}\right)}}.
\end{equation}
The minimal area surface $\Sigma_0$ can be found by solving the equation of motion arising from the above action
\begin{align}
\frac{dx}{dr}= \pm \frac{r_c^2}{ r^4 \sqrt{\left(1- \frac{r_c^4}{r^4}\right)\left(1-\frac{r_H^d}{r^d}\right)}},
\end{align}
where, $r_c$ is an integral of motion and $r=r_c$ represents the point of closest approach of the extremal surface.\footnote{It should be noted that $r_c$ is an observable dependent quantity and hence $r_c$ for the spatial rectangular Wilson loop is not the same as $r_c$ for equal-time two-point function.} Each surface, just like the geodesic case, has two branches, joined smoothly at $(r=r_c, x=0)$ and $r_c$ can be determined using the boundary conditions:
\begin{equation}
x(\infty)=\pm\frac{l}{2}.
\end{equation}
That leads to
\begin{align}
\frac{l}{2}=&\int_{r_c}^{\infty}\frac{r_c^2 dr}{r^4 \sqrt{\left(1- \frac{r_c^4}{r^4}\right)}} \left(1-\frac{r_H^d}{r^d}\right)^{-1/2}\nonumber\\
=& \frac{1}{r_c}\int_{0}^{1}\frac{u^2 du}{ \sqrt{1- u^4}} \left(1-\frac{r_H^d}{r_c^d}u^d\right)^{-1/2}.
\end{align}
Again, analogous to the two-point function case, we can develop a systematic expansion, 
\begin{align}
l=\frac{1}{2 r_c}\sum_{n=0}^\infty\frac{\Gamma\left[\frac{1}{2}+n\right]\Gamma\left[\frac{1}{4}(3+nd)\right]}{ \Gamma[1+n]\Gamma\left[\frac{1}{4}(5+nd)\right]}\left(\frac{r_H}{r_c}\right)^{nd}.\label{wlrc}
\end{align}
For large $n$ the series goes as $\sim \frac{1}{n}(r_H/r_c)^{n d}$ and the series converges as long as $r_H/r_c<1$. Again it can be shown \cite{Hubeny:2012ry} that at any finite temperature $r_c>r_H$ and hence the sum (\ref{wlrc}) is well-defined. Next we will calculate the on-shell action
\begin{align}
S_{NG}=&\frac{2 L}{2 \pi \alpha '} \int_{r_c}^{\infty}\frac{ dr}{ \sqrt{\left(1- \frac{r_c^4}{r^4}\right)}} \left(1-\frac{r_H^d}{r^d}\right)^{-1/2}. 
\end{align}      
The factor of 2 appears because of the two branches of the extremal surface. The on-shell $S_{NG}$ is divergent;\footnote{Note that the divergent piece is independent of both temperature and dimension.} however, it can be renormalized by introducing a UV-cutoff $r_b$ and then subtracting the boundary term $r_b L/(\pi \alpha ')$
\begin{equation}
S_{NG;ren}=S_{NG}-\frac{r_b L}{ \pi \alpha '}.
\end{equation}
Now a series expansion for $S_{NG;ren}$ can be obtained
\begin{align}
S_{NG;ren}=\frac{L r_c}{\pi \alpha'}\left[-\frac{\sqrt{2}\pi^{3/2}}{\Gamma(\frac{1}{4})^2}+\frac{1}{4}\sum_{n=1}^\infty \frac{\Gamma\left[\frac{1}{2}+n\right]\Gamma\left[\frac{1}{4}(-1+nd)\right]}{ \Gamma[1+n]\Gamma\left[\frac{1}{4}(1+nd)\right]}\left(\frac{r_H}{r_c}\right)^{nd}\right].\label{sngren}
\end{align}
Again it can be shown that the series (\ref{sngren}) converges for $r_H/r_c<1$. In principle, the rest of the calculation is straight forward. We have to solve equation (\ref{wlrc}) for $r_c$ and then use that $r_c$ in equation (\ref{sngren}) to get $S_{NG;ren}$. The expectation value of the renormalized Wilson loop is then given by,
\begin{equation}
\langle W(\C)\rangle=e^{- S_{NG;ren}}.
\end{equation}
%%%%%%%%%%%%%%%%%%%%%%%%%%%%%%%%%%%%%%%%%%%%%%%%%%%%%%%%%%%%%%%%%%%%%%%%%

\subsection{Rectangular Wilson loop: Low temperature limit}
At low temperature (in this context, low temperature means $T l\ll 1$), the calculation is similar to the two-point function case (see appendix \ref{lte}). The leading contributions come from the boundary and hence we can solve equation (\ref{wlrc}) for $r_c$ order by order, leading to finite temperature corrections to the expectation value of the Wilson loop. Solving (\ref{wlrc}), at first order in $(l r_H)^d$, we obtain
\begin{align}
r_c=\frac{1}{l}\left(\frac{2 \sqrt{2} \pi ^{3/2}}{ \Gamma \left(\frac{1}{4}\right)^2} \right)\left[1+\frac{2^{-\frac{3 d}{2}-\frac{7}{2}} \pi ^{-\frac{3 d}{2}-1} \Gamma \left(\frac{1}{4}\right)^{2 d+2} \Gamma \left(\frac{d+3}{4}\right) }{\Gamma \left(\frac{d+5}{4}\right)}\left(l r_H\right){}^d + \O(l r_H)^{2d}\right] .
\end{align}
Equation (\ref{sngren}) then tells us
\begin{align}
S_{NG;ren}=-\frac{4 \pi^2  }{\alpha' \Gamma(\frac{1}{4})^4}\left(\frac{L}{l}\right)\left[1+ \D_1 (r_H l)^d +\O(r_H l)^{2d}\right]
\end{align} 
where $\D_1$ is a numerical constant given by
\begin{equation}
\D_1= -\frac{2^{-\frac{3 d}{2}-\frac{9}{2}} \pi ^{-\frac{3 d}{2}-1} \Gamma \left(\frac{1}{4}\right)^{2 d+2} \Gamma \left(\frac{d}{4}-\frac{1}{4}\right)}{\Gamma \left(\frac{d}{4}+\frac{5}{4}\right)}.
\end{equation}
Expectation value of the renormalized Wilson loop is then given by $\langle W(\C)\rangle=e^{- S_{NG;ren}}$. Constant $\D_1$ is negative indicating a decrease in the expectation value as the system is heated. Replacing $r_H$ by corresponding temperature (\ref{temperature}), for the $d$-dimensional boundary theory we obtain
\begin{equation}\label{lowtempwl}
S_{NG;ren}=-\frac{4 \pi^2 }{ \Gamma(\frac{1}{4})^4}\left(\frac{R^2}{\alpha'}\right)\left(\frac{L}{l}\right)\left[1+ \D_1 \left(\frac{4 \pi T l}{d}\right)^d +\O\left(\frac{4 \pi T l}{d}\right)^{2d}\right],
\end{equation}
where, we have restored AdS radius $R$ by dimensional analysis. Particularly, for $d=4$ at low temperature limit, we get
\begin{equation}
S_{NG;ren}=-\frac{4 \pi^2 }{ \Gamma(\frac{1}{4})^4}\left(\frac{R^2}{\alpha'}\right)\left(\frac{L}{l}\right)\left[1-\frac{\Gamma(1/4)^8}{320 \pi^6} ( \pi T l)^4 +\O(\pi T l)^{8}\right].
\end{equation}
%%%%%%%%%%%%%%%%%%%%%%%%%%%%%%%%%%%%%%%%%%%%%%%%%%%%%%%%%%%%%%%%%%%%%%%%%%%
\subsection{Rectangular Wilson loop: High temperature limit}
At high temperature ($r_H l\gg 1$), just like the geodesic case, $r_c$ approaches $r_H$ and the leading contribution comes from the near horizon region. We can rewrite equation (\ref{sngren}) in a way so that we can take the limit $r_c\rightarrow r_H$
\begin{align}
S_{NG;ren}=\frac{L r_c }{\pi \alpha'}\left[-\frac{2\sqrt{2}\pi^{3/2}}{\Gamma(\frac{1}{4})^2}+\frac{l r_c}{2}+\frac{1}{2}\sum_{n=1}^\infty\frac{1}{nd-1}\frac{\Gamma\left[\frac{1}{2}+n\right]\Gamma\left[\frac{1}{4}(3+nd)\right]}{ \Gamma[1+n]\Gamma\left[\frac{1}{4}(5+nd)\right]}\left(\frac{r_H}{r_c}\right)^{nd}\right].
\end{align}
It is easy to check that the infinite series in the last equation converges for $r_c\geq r_H$ and we can safely take the limit  $r_c\rightarrow r_H$
\begin{align}
S_{NG;ren}\approx\frac{L r_H }{\pi \alpha'}\left[-\frac{2\sqrt{2}\pi^{3/2}}{\Gamma(\frac{1}{4})^2}+\frac{l r_H}{2}+\frac{1}{2}\sum_{n=1}^\infty\frac{1}{nd-1}\frac{\Gamma\left[\frac{1}{2}+n\right]\Gamma\left[\frac{1}{4}(3+nd)\right]}{ \Gamma[1+n]\Gamma\left[\frac{1}{4}(5+nd)\right]}\right].
\end{align}
Therefore, at high temperature the expectation value of the Wilson loop is approximately given by 
\begin{equation}
\langle W(\C)\rangle\approx e^{-  \frac{\W_d L r_H }{\pi \alpha'}}~e^{-\frac{A r_H^2 }{2\pi \alpha'}}
\end{equation}
where, $A=lL$ is the area of the Wilson loop and
\begin{equation}
\W_d=-\frac{2\sqrt{2}\pi^{3/2}}{\Gamma(\frac{1}{4})^2}+\frac{1}{2}\sum_{n=1}^\infty\frac{1}{nd-1}\frac{\Gamma\left[\frac{1}{2}+n\right]\Gamma\left[\frac{1}{4}(3+nd)\right]}{ \Gamma[1+n]\Gamma\left[\frac{1}{4}(5+nd)\right]}.
\end{equation}
In terms of temperature $T$, the asymptotic behavior of the expectation value of the Wilson loop in $d-$dimensions is given by (after restoring AdS radius $R$)
\begin{equation}\label{hightempwl}
\langle W(\C)\rangle\approx \exp\left[- \left(\frac{R^2}{\alpha'}\right)\left( \frac{8 \pi A T^2 }{d^2}+\frac{4 \W_d L T }{d}\right)\right].
\end{equation}
At large temperature, the term proportional to the area in $S_{NG;ren}$ dominates indicating an area law which is consistent with \cite{Brandhuber:1998er}. The area term comes from the near horizon part of the extremal surface and could be understood easily by making an argument similar to the geodesic case (Figure (\ref{approxgeodesic}) could be think of as a section of the extremal surface). The term independent of $l$ receives contributions from the full bulk geometry.

At high temperature $(Tl\gg 1)$, similar to the two-point function case, $r_c$ approaches $r_H$ exponentially fast (see appendix \ref{subleading})
\begin{equation}
r_c=r_H\left(1+ \E_{wl} ~e^{-\sqrt{d}~l r_H}+...\right),
\end{equation}
where $\E_{wl}$ is a numerical constant given by,
\begin{equation}
\E_{wl}=\frac{1}{d}\exp\left[\sqrt{d}\left\{\frac{2\sqrt{2}\pi^{3/2}}{ \Gamma(\frac{1}{4})^2}+ \frac{1}{2 }\sum_{n=1}^\infty \left(\frac{\Gamma\left[\frac{1}{2}+n\right]\Gamma\left[\frac{1}{4}(3+nd)\right]}{ \Gamma[1+n]\Gamma\left[\frac{1}{4}(5+nd)\right]}-\frac{2}{\sqrt{d} n}\right)\right\}\right].
\end{equation} 
Using this expression for $r_c$, it has been shown in appendix \ref{subleading} that the correction to the high temperature result (\ref{hightempwl}) is very small
\begin{align}\label{highwl}
S_{NG;ren}=\left(\frac{R^2}{\alpha'}\right)\left(\frac{8 \pi A T^2 }{d^2}+\frac{4 L T \W_d}{d}-\frac{4 L T }{d^{3/2}}\E_{wl} ~e^{-\frac{4}{ \sqrt{d}}\pi l T}+...\right)
\end{align}
where the dots represent the higher order correction terms.

%%%%%%%%%%%%%%%%%%%%%%%%%%%%%%%%%%%%%%%%%%%%%%%%%%%%%%%%%%%%%%%%%%%%%%%%%%%%%%%%%%%%%%

\section{Entanglement entropy}\label{secee}
The entanglement entropy in quantum field theories or quantum many body systems is another important non-local quantity that can be employed to probe quantum properties of the system. Consider a quantum field theory with many degrees of freedom at zero temperature. Let us assume that the system is described by the pure ground state $|\Psi\rangle$, which does not have any degeneracy. The density matrix of the state is
\begin{equation}
\rho_{tot}=|\Psi\rangle\langle\Psi| 
\end{equation}
and hence the von Neumann entropy of the total system, defined as $S_{tot}=-\tr (\rho_{tot}\ln \rho_{tot})=0$, does not contain any useful information. On the other hand, the entanglement entropy is non-vanishing even at zero temperature and several aspects of quantum many body physics can be understood by studying the entanglement entropy associated with volumes of different shapes and sizes.

\begin{figure}[!ht]
\begin{center}
\subfigure[]{\includegraphics[angle=0, trim = 10mm 0mm 10mm 20mm, clip, width=0.35\textwidth]{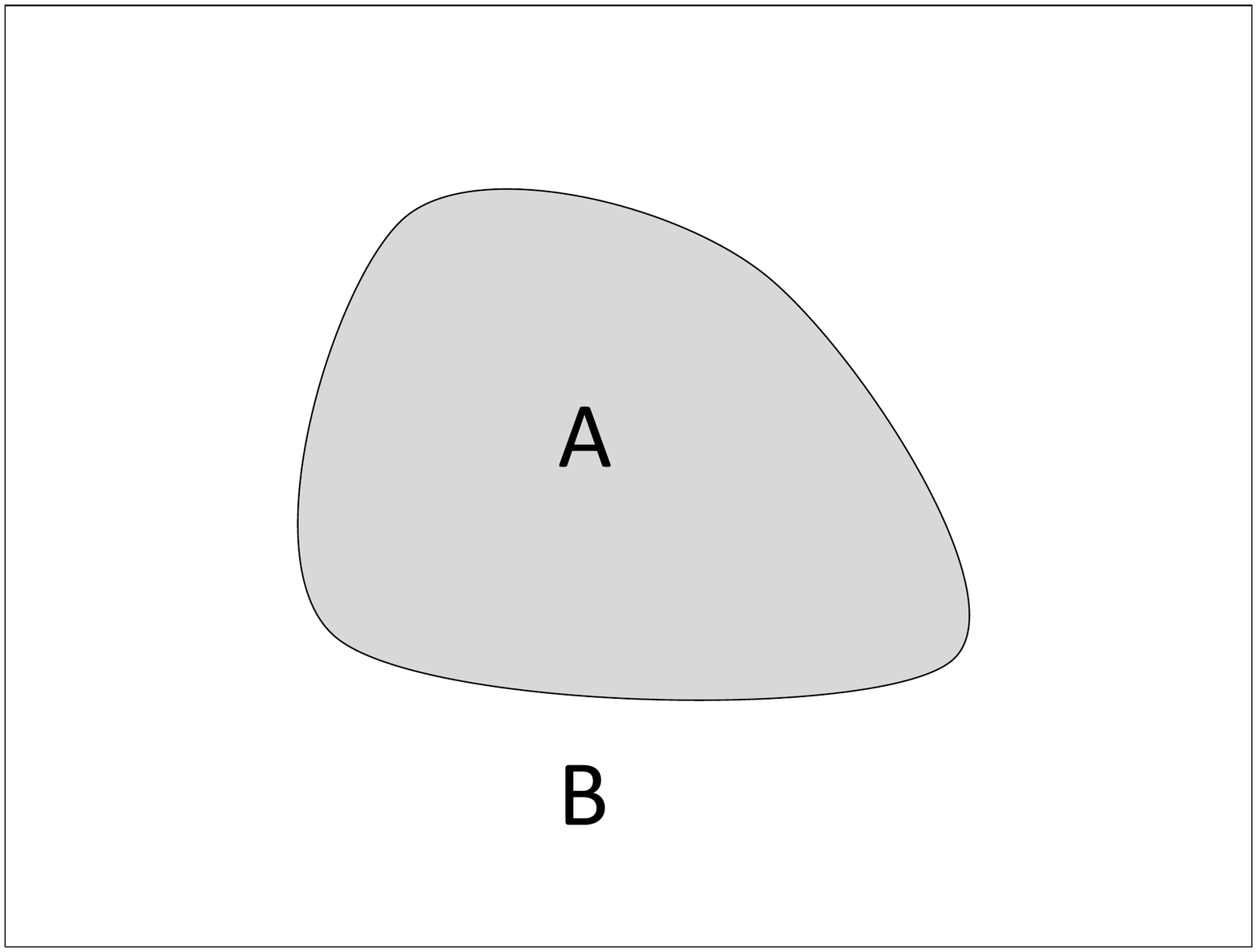} \label{ee1}} 
\subfigure[] {\includegraphics[angle=0, trim = 20mm 45mm 15mm 20mm, clip,width=0.60\textwidth]{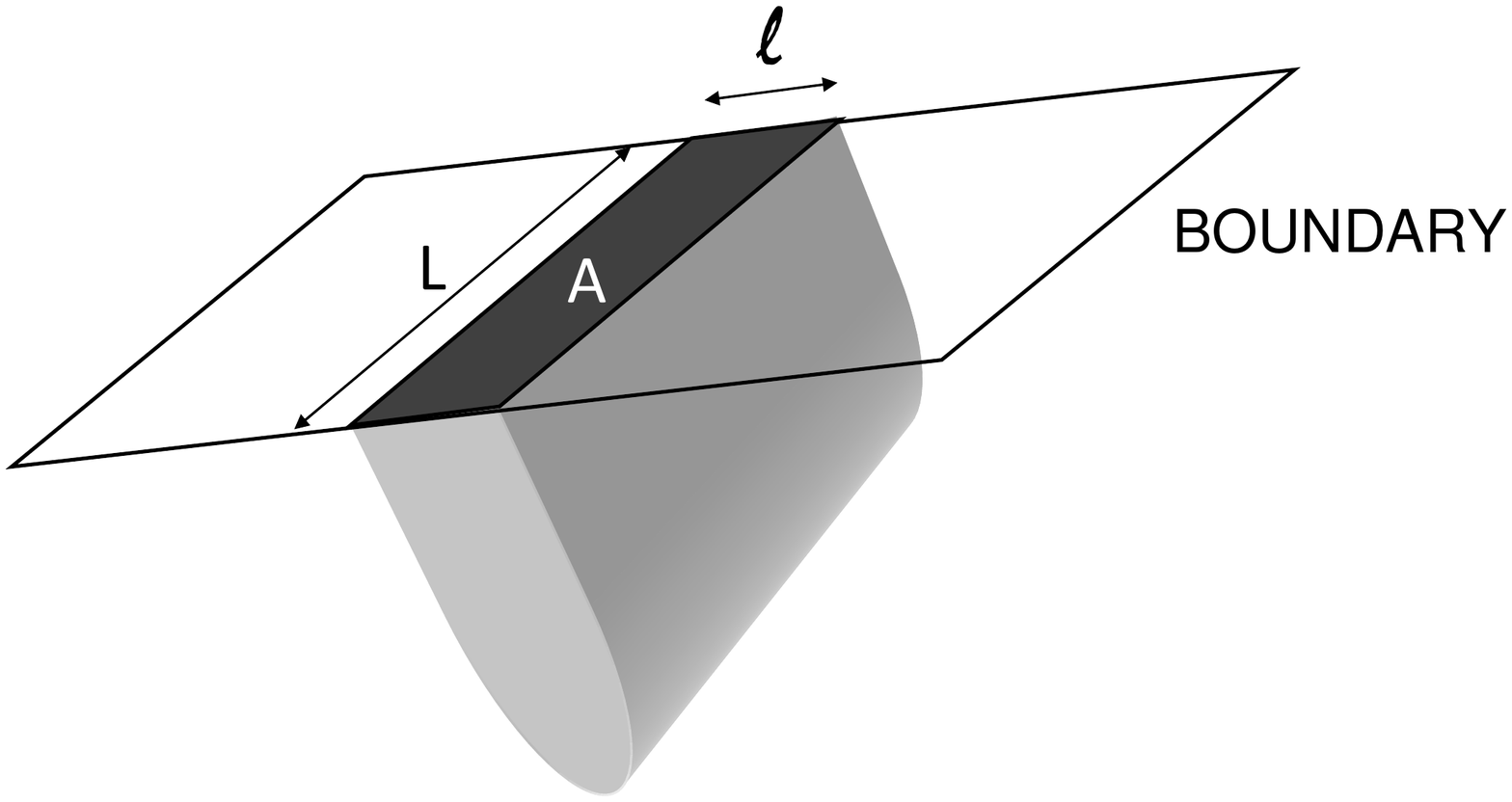} \label{holo}} 
\caption{\small (a) The total system can be divided into two subsystems A and B; the entanglement entropy $S_A$ measures the amount of information loss because of smearing out in region B. (b) A schematic diagram of the rectangular geometry and the corresponding extremal surface used for the calculations of spatial Wilson loops and the entanglement entropy.}
\label{wlee}
\end{center}
\end{figure}

Now consider dividing the total system into two subsystems A and B and imagine an observer who has access only to the subsystem A (see figure \ref{ee1}). The total Hilbert space is a direct product of two spaces $\H_{tot}=\H_A \otimes\H_B$. The observer who is restricted to A will describe the total system by the reduced density matrix
\begin{equation}
\rho_A=\tr_B \rho_{tot}, 
\end{equation}
where, degrees of freedom inside the subsystem B have been traced out. The entanglement entropy of the subsystem A is now defined as the von Neumann entropy of the reduced density matrix $\rho_A$
\begin{equation}
S_A=-\tr_A(\rho_A \ln \rho_A).
\end{equation}
The entanglement entropy $S_A$ describes the amount of information loss because of smearing out in region B and provides us with a convenient way to measure how the subsystems A and B are correlated with each other. At finite temperature, the entanglement entropy is defined in the same way but with the thermal density matrix $\rho=e^{-H/T}$, where $H$ is the total Hamiltonian.   

A precise prescription for computing entanglement entropy for strongly coupled field theories with AdS duals was proposed in \cite{Ryu:2006bv} and later generalized in \cite{Hubeny:2007xt} (for a good review see \cite{Nishioka:2009un}). According to the proposal, the entanglement entropy $S_A$ is given by
\begin{equation}\label{ee}
S_A=\frac{\text{Area}\left(\gamma_A\right)}{4 G_N^{(d+1)}}, 
\end{equation}
where, $G_N^{(d+1)}$ is the $(d+1)$-dimensional Newton's constant. $\gamma_A$ is the $(d-1)$-dimensional minimal area surface in the bulk whose boundary is given by the boundary of the region A: $\partial \gamma_A=\partial A$. The area of the surface $\gamma_A$ is denoted by $\text{Area}\left(\gamma_A\right)$.

From this prescription it is very clear that for $d=2$ and $3$, the entanglement entropy calculation is the same as the equal time two-point function calculation and the spatial Wilson loop calculation respectively. Therefore, only for $d\ge 4$, the entanglement entropy contains non-trivial information.
%%%%%%%%%%%%%%%%%%%%%%%%%%%%%%%%%%%%%%%%%%%%%%%%%%%%%%%%%%%%%%%%%%%%%%%%%%%%%%%%%%%%%%
\subsection{Entanglement entropy: Finite temperature expansion}
In this section we will compute the entanglement entropy for a strip (see figure \ref{holo} for a schematic diagram) specified by 
\begin{equation}
x\equiv x^1 \in \left[-\frac{l}{2},\frac{l}{2}\right],~  x^i\in \left[-\frac{L}{2},\frac{L}{2}\right], i=2,...,d-1
\end{equation}
with $L \rightarrow \infty$. Extremal surface is translationally invariant along $x^i, i=2,...,d-1$ and the profile of the surface in the bulk is $x(r)$. Area of this surface is given by
\begin{equation}
A= L^{d-2}\int dr r^{d-2}\sqrt{r^2 x'^2+ \frac{1}{r^2\left(1-\frac{r_H^d}{r^d}\right)}}.
\end{equation}
This action leads to the equation of motion
\begin{align}
\frac{dx}{dr}= \pm \frac{r_c^{d-1}}{ r^{d+1} \sqrt{\left(1- \frac{r_c^{2d-2}}{r^{2d-2}}\right)\left(1-\frac{r_H^d}{r^d}\right)}},
\end{align}
where, $r_c$ is an integral of motion and $r=r_c$ represents the point of closest approach of the extremal surface (again it should be noted that $r_c$ is an observable dependent quantity). Similar to the previous cases, each surface has two branches, joined smoothly at $(r=r_c, x=0)$ and $r_c$ can be determined using the boundary conditions:
\begin{equation}
x(\infty)=\pm\frac{l}{2}.
\end{equation}
That leads to
\begin{align}
\frac{l}{2}=&\int_{r_c}^{\infty}\frac{r_c^{d-1} dr}{r^{d+1} \sqrt{\left(1- \frac{r_c^{2d-2}}{r^{2d-2}}\right)}} \left(1-\frac{r_H^d}{r^d}\right)^{-1/2}\nonumber\\
=& \frac{1}{r_c}\int_{0}^{1}\frac{u^{d-1} du}{ \sqrt{1- u^{2d-2}}} \left(1-\frac{r_H^d}{r_c^d}u^d\right)^{-1/2}.
\end{align}
By now it is obvious that we will do an expansion
\begin{align}
l=\frac{2}{ r_c}\sum_{n=0}^\infty\left(\frac{1}{1+nd}\right)\frac{\Gamma\left[\frac{1}{2}+n\right]\Gamma \left[\frac{d (n+1)}{2 (d-1)}\right]}{ \Gamma[1+n]\Gamma \left[\frac{d n+1}{2 (d-1)}\right]}\left(\frac{r_H}{r_c}\right)^{nd}.\label{eerc}
\end{align}
Again it can be shown that the series converges for $r_c> r_H$. The area of the extremal surface is given by,
\begin{align}
A=2 L^{d-2}\int_{r_c}^{\infty}\frac{r^{d-3} dr}{ \sqrt{\left(1- \frac{r_c^{2d-2}}{r^{2d-2}}\right)}} \left(1-\frac{r_H^d}{r^d}\right)^{-1/2}
\end{align}
This area is infinite indicating that the entanglement entropy has a divergence. In a field theory the entanglement entropy is always divergent because there are too many degrees of freedom. Therefore, we can write 
\begin{equation}
A=A_{div}+A_{finite}. 
\end{equation}
The divergent piece is temperature independent and hence easy to compute. We will introduce an infrared cut off $r_b$ which corresponds to the ultraviolet cut off $a=1/r_b$ (or a lattice spacing) of the boundary theory\footnote{We are working with AdS radius $R=1$. Restoring $R$, the lattice spacing is given by $a=\frac{R^2}{r_b}$.}
\begin{equation}
A_{div}=\frac{2}{d-2}L^{d-2}{r_b}^{d-2}= \frac{2}{d-2}\left(\frac{L}{a}\right)^{d-2}\qquad d\neq 2.
\end{equation}
The divergence is proportional to the area of the boundary of A which is expected since the entanglement between A and B is strongest at the boundary $\partial A$. This area law behavior of the divergent piece is well understood from field theory computations \cite{Bombelli:1986rw, Srednicki:1993im, Casini:2003ix, Plenio:2004he, Cramer:2005mx, Das:2005ah}.

Analogous to the previous cases, we can do an expansion $(d\neq 2)$ for the finite part of the area\footnote{We will not consider the entanglement entropy for $d=2$ case because the calculations are exactly the same as the two-point function case and hence has been studied analytically. In $d=2$, the entanglement entropy has a logarithmic divergence \cite{Holzhey:1994we, Calabrese:2004eu}, indicating a violation of the simple area law.}
\begin{align}
A_{finite}=&2 L^{d-2}r_c^{d-2}\int_{r_c/r_b}^{1}\frac{ du}{u^{d-1} \sqrt{1- u^{2d-2}}} \left(1-\frac{r_H^d}{r_c^d}u^d\right)^{-1/2}-\frac{2}{d-2}L^{d-2}{r_b}^{d-2}\nonumber\\
=&2 L^{d-2}r_c^{d-2} \left[\frac{\sqrt{\pi } \Gamma \left(-\frac{d-2}{2 (d-1)}\right)}{2 (d-1) \Gamma \left(\frac{1}{2 (d-1)}\right)}+ \sum_{n=1}^\infty\left(\frac{1}{2(d-1)}\right)\frac{\Gamma\left[\frac{1}{2}+n\right]\Gamma \left[\frac{d (n-1)+2}{2 d-2}\right]}{ \Gamma[1+n]\Gamma \left[\frac{d n+1}{2 (d-1)}\right]}\left(\frac{r_H}{r_c}\right)^{nd}\right].\label{aren}
\end{align}
It can be shown that this series converges for $r_c>r_H$. Now, the rest of the procedure is simple and familiar. We have to solve equation (\ref{eerc}) for $r_c$ and then we can calculate area by using equation (\ref{aren}). Entanglement entropy of the rectangular strip can be computed using the relation (\ref{ee}). In practice, this procedure can not be performed analytically at finite temperature. However, we can extract low and high temperature behavior of the entanglement entropy from equations (\ref{eerc},\ref{aren}).

%%%%%%%%%%%%%%%%%%%%%%%%%%%%%%%%%%%%%%%%%%%%%%%%%%%%%%%%%%%%%
\subsection{Entanglement entropy: Low temperature limit}
The temperature now should be measured with respect to $\sim 1/l$; therefore, low temperature means $Tl\ll 1$. At low temperature, $r_c\gg r_H$ and the leading contributions to the area come from the boundary which is still AdS. Therefore we should expect the zero temperature entanglement entropy as the leading term. Finite temperature corrections correspond to the deviation of the bulk geometry from pure AdS. At low temperature, the extremal surface is restricted to be near the boundary region and hence the deviation is small and can be computed perturbatively. At low temperature limit ($r_H l\ll 1$), equation (\ref{eerc}) can be solved for $r_c$ and at first order in $(r_H l)^d$, we obtain
\begin{align}
r_c=\frac{2 \sqrt{\pi}\Gamma\left[\frac{d}{2(d-1)}\right]}{l~\Gamma\left[\frac{1}{2(d-1)}\right]} \left[1+ \frac{1}{2(d+1)}\frac{2^{\frac{1}{d-1}-d} \Gamma \left(1+\frac{1}{2 (d-1)}\right) \Gamma \left(\frac{1}{2 (d-1)}\right)^{d+1}}{\pi^{\frac{d+1}{2}}  \Gamma \left(\frac{1}{2}+\frac{1}{d-1}\right)\Gamma\left(\frac{d}{2(d-1)}\right)^d}\left(r_H l\right)^d+\O\left(r_H l\right)^{2d}\right]
\end{align}
Now using equation (\ref{aren}), at first order in $(r_H l)^d$, we get (the calculation is similar to the two-point function calculation explained in appendix \ref{lte})
\begin{align}
A_{finite}= \S_0 \left(\frac{L}{l}\right)^{d-2}\left[1+ \S_1 (r_H l)^d+ \O(r_H l)^{2d}\right]
\end{align}
where, numerical constants $\S_0, \S_1$ are given by
\begin{align}
\S_0=& \frac{2^{d-2} \pi ^{\frac{d-1}{2}} \Gamma \left(-\frac{d-2}{2 (d-1)}\right) }{(d-1) \Gamma \left(\frac{1}{2 (d-1)}\right)} \left(\frac{\Gamma \left(\frac{d}{2 (d-1)}\right)}{\Gamma \left(\frac{1}{2 (d-1)}\right)}\right)^{d-2}\\
\S_1=& \frac{\Gamma \left(\frac{1}{2 (d-1)}\right)^{d+1}}{\Gamma \left(\frac{d}{2(d-1)}\right)^d\Gamma \left(\frac{1}{2}+\frac{1}{d-1}\right)}2^{-d-1} \pi ^{-\frac{d}{2}} \left(\frac{\Gamma \left(\frac{1}{d-1}\right) }{\Gamma \left(-\frac{d-2}{2 (d-1)}\right)}+\frac{2^{\frac{1}{d-1}} (d-2) \Gamma \left(1+\frac{1}{2 (d-1)}\right) }{\sqrt{\pi } (d+1)}\right)
\end{align}
Therefore, following equation (\ref{ee}), after restoring AdS radius $R$, the entanglement entropy of the rectangular strip for the $d$-dimensional boundary theory at low temperature ($Tl\ll1$) is given by,
\begin{equation}\label{eel}
S_A=\frac{R^{d-1}}{4 G_N^{(d+1)}}\left[\frac{2}{d-2}\left(\frac{L}{a}\right)^{d-2}+ \S_0 \left(\frac{L}{l}\right)^{d-2}\left\{1+ \S_1 \left(\frac{4 \pi T l}{d}\right)^d+ \O\left(\frac{4 \pi T l}{d}\right)^{2d}\right\}\right].
\end{equation} 
Note that in the limit $T\rightarrow 0$, we recover the well known results of \cite{Ryu:2006ef}. Particularly for $d=4$ we have,
\begin{equation}
S_A=\frac{R^{d-1}}{4 G_N^{(d+1)}}\left[\left(\frac{L}{a}\right)^2 -0.32 \left(\frac{L}{l}\right)^{2}\left\{1-(1.764) (\pi T l)^4+ \O(\pi T l)^{8}\right\}\right].
\end{equation}
%%%%%%%%%%%%%%%%%%%%%%%%%%%%%%%%%%%%%%%%%%%%%%%%%%%%%%%%%%%%%
\subsection{Entanglement entropy: High temperature limit}
At high temperature (i.e. $T l \gg 1$), it is not very difficult to find out the asymptotic behavior of the entanglement entropy. However, it is more difficult to do a systematic expansion. At high temperature, the extremal surface tends to wrap a part of the horizon and the leading contribution comes from this near horizon part of the surface. For the subleading terms, the full bulk geometry contributes and they are more interesting.

At very high temperature ($r_H l\gg 1$), $r_c$ approaches $r_H$. By now we know how to determine the high temperature asymptotic behavior; we will rewrite equation (\ref{aren}) in a way that allows us to take the limit  $r_c\rightarrow r_H$ without encountering any divergence.
\begin{align}
A_{finite}=&2 L^{d-2}r_c^{d-2} \left[\frac{\sqrt{\pi } \Gamma \left(-\frac{d-2}{2 (d-1)}\right)}{2 (d-1) \Gamma \left(\frac{1}{2(d-1)}\right)}\right.\nonumber\\&\left.+ \sum_{n=1}^\infty\frac{1}{1+nd}\left(1+\frac{d-1}{d(n-1)+2}\right)\frac{\Gamma\left[\frac{1}{2}+n\right]\Gamma \left[\frac{d (n+1)}{2 d-2}\right]}{ \Gamma[1+n]\Gamma \left[\frac{d n+1}{2 (d-1)}\right]}\left(\frac{r_H}{r_c}\right)^{nd}\right]\nonumber\\
=&2 L^{d-2}r_c^{d-2} \left[\frac{l r_c}{2}-\frac{\sqrt{\pi }(d-1) \Gamma \left(\frac{d}{2 (d-1)}\right)}{(d-2)\Gamma \left(\frac{1}{2(d-1)}\right)}\right.\nonumber\\
&+\left.\sum_{n=1}^\infty\left(\frac{1}{1+nd}\right)\left(\frac{d-1}{d(n-1)+2}\right)\frac{\Gamma\left[\frac{1}{2}+n\right]\Gamma \left[\frac{d (n+1)}{2 d-2}\right]}{ \Gamma[1+n]\Gamma \left[\frac{d n+1}{2 (d-1)}\right]}\left(\frac{r_H}{r_c}\right)^{nd}\right].\label{highA}
\end{align}
The infinite series in the last equation for large $n$ goes as $\sim \frac{1}{n^2} (r_H/r_c)^{nd}$ and thus the limit $r_c\rightarrow r_H$ exists. At high temperature $r_c\sim r_H$ and  the leading behavior can be determined by taking the limit $r_c\rightarrow r_H$ in the last equation
\begin{align}\label{Shigh}
A_{finite}\approx l L^{d-2}r_H^{d-1}\left[1+ \left(\frac{1}{l r_H}\right)\S_{high}\right]
\end{align}
where, $\S_{high}$ is another numerical constant given by
\begin{align}
\S_{high}=&2\left[-\frac{\sqrt{\pi }(d-1) \Gamma \left(\frac{d}{2 (d-1)}\right)}{(d-2)\Gamma \left(\frac{1}{2(d-1)}\right)}+\sum_{n=1}^\infty\left(\frac{1}{1+nd}\right)\left(\frac{d-1}{d(n-1)+2}\right)\frac{\Gamma\left[\frac{1}{2}+n\right]\Gamma \left[\frac{d (n+1)}{2 d-2}\right]}{ \Gamma[1+n]\Gamma \left[\frac{d n+1}{2 (d-1)}\right]}\right].\label{constantshigh}
\end{align}
Hence, the entanglement entropy of the rectangular strip for the $d$-dimensional boundary theory at high temperature is given by,
\begin{equation}\label{highee}
S_A\approx\frac{R^{d-1}}{4 G_N^{(d+1)}} \left[\frac{2}{d-2}\left(\frac{L}{a}\right)^{d-2}+V \left(\frac{4 \pi T}{d}\right)^{d-1}\left\{1+ \left(\frac{d}{4 \pi T l }\right)\S_{high}\right\}\right]
\end{equation}
where $V=l L^{d-2}$ is the volume of the rectangular strip and $R$ is the AdS radius.

The divergent part of the entanglement entropy is temperature independent and thus it does not contain any new information. The leading finite piece in equation (\ref{highee}) is proportional to the volume of the rectangular strip and it is just the thermal entropy of the region A.  The extrinsic nature of the  leading term at high temperature can be understood very easily by looking at the extremal surface for $r_H l\gg 1$. In this limit, the extremal surface tends to wrap a part of the horizon and the actual U-shaped surface can be approximated by a surface that consists of $x=-l/2, r=r_H, x=l/2$ (one can think of figure (\ref{approxgeodesic}) as a section of the extremal surface). At high temperature limit, the most dominant contribution to the area of the extremal surface comes from the near horizon part which can be guessed from this approximate surface. The area of the near horizon part of the approximate surface is $A\sim r_H^{d-1} V$. Therefore, it is expected that the leading term goes as $S_A\sim V T^{d-1}$. On the other hand, the other term $\sim (LT)^{d-2}$ is more interesting. This term corresponds to the entanglement between the region A and the outside and it is proportional to the area of the boundary $\partial A$ of A because the entanglement is strongest at the boundary. One can guess the functional form of this term from the approximate surface. However, the value of the numerical constant $\S_{high}$ obtained from the approximate surface is inaccurate. Although the actual extremal surface approaches the approximate one as $r_H l$ is increased,  it can be shown that the area of this approximate surface is always greater than the actual extremal surface \cite{Faraggi:2007fu} and even in the limit $r_H l \rightarrow \infty$, two do not coincide.

So far we have used the fact that at high temperature, $r_c\sim r_H$. \cite{Hubeny:2012ry} shows that the extremal surface approaches the horizon but it always stays finite distance above the horizon. It is an interesting exercise to see exactly how fast  $r_c$ approaches $r_H$. As shown in appendix (\ref{subleading}), in the limit $r_H l \gg 1$
\begin{equation}\label{rcee}
r_c=r_H\left(1+\E_{ent} ~e^{-\sqrt{\frac{d(d-1)}{2}}l r_H}+...\right),
\end{equation}
where $\E_{ent}$ is a constant given by,
\begin{align}
\E_{ent}=&\frac{1}{d}\exp\left[\sqrt{\frac{d(d-1)}{2}}\left\{\frac{2\sqrt{\pi } \Gamma \left(\frac{d}{2 (d-1)}\right)}{\Gamma \left(\frac{1}{2 (d-1)}\right)}\right.\right.\nonumber\\
&\left.\left.+ 2\sum_{n=1}^\infty\left\{\left(\frac{1}{1+nd}\right)\frac{\Gamma\left[\frac{1}{2}+n\right]\Gamma \left[\frac{d (n+1)}{2 (d-1)}\right]}{ \Gamma[1+n]\Gamma \left[\frac{d n+1}{2 (d-1)}\right]}-\frac{1}{\sqrt{2} \sqrt{(d-1) d}~ n}\right\}\right\}\right].
\end{align}
Using equation (\ref{rcee}), we can calculate the next order correction to the high temperature entanglement entropy (for details see appendix (\ref{subleading})). The subleading term is exponentially suppressed
\begin{align}
S_A=S_{div}+&\frac{R^{d-1}}{4 G_N^{(d+1)}} \left(\frac{4 \pi }{d}\right)^{d-1}\left[V T^{d-1}+ \left(\frac{ \S_{high} d}{8 \pi }\right)A~ T^{d-2}\right.\nonumber\\
&\left.-\left(\frac{ \E_{ent}}{8 \pi }\right) \sqrt{2d(d-1)}~ A~ T^{d-2}~\exp\left\{-\sqrt{(d-1)/(2d)}~4 \pi T l \right\}+...\right]\label{eeh}
\end{align}
where $A$ is the area $A=2 L^{d-2}$ and the dots represent the higher order correction terms. 
%%%%%%%%%%%%%%%%%%%%%%%%%%%%%%%%%%%%%%%%%%%%%%%%%%%%%%%%%%%%%%%%%%%%%%%%%%%%%%%%%%%%%%
\subsection{Entanglement entropy of a generic region at high temperature}
The calculation of the entanglement entropy of an infinite rectangular strip suggests that the general form of the finite part of the high temperature answer does not particularly depend on the shape. One expects that the finite part of the entanglement entropy of a region $A$ for a $d-$dimensional ($d>2$) boundary theory with AdS-dual should be given by
\begin{align}
S_{A;finite}=c_0 &\left[ T^{d-1}\text{Volume}(A)+c_1 ~T^{d-2}\text{Area}(\partial A)\right]+ \text{sub-leading terms},
\end{align}
provided the temperature $T\gg1/l$, where $l$ is the smallest length scale of the region $A$.\footnote{One can make a similar argument for the spatial Wilson loop.} $c_0$ is a constant that depends on the actual theory and $c_1$ is a constant that depends on the shape of the region $A$. The first term represents the thermal entropy of the region $A$. The second term is proportional to the area of the boundary of $A$ and corresponds to the entanglement between region $A$ and the outside.
%%%%%%%%%%%%%%%%%%%%%%%%%%%%%%%%%%%%%%%%%%%%%%%%%%%%%%%%%%%%%%%%%%%%%%%%%%%%%%%%%%%%%
\section{Non-relativistic theories with hyperscaling violation}\label{hvb}
We can generalize the techniques that we have developed in the previous sections to study a broader class of strongly coupled large-N field theories that have the following properties
\begin{eqnarray}
t \to \lambda^z t \ , \quad r \to \lambda r \ , \quad x \to \lambda x \ , \quad ds^2 \to \lambda^{2\theta/(d-1)} ds^2 \ 
\end{eqnarray}
and are dual to
\begin{eqnarray}
ds^2 = \frac{1}{r^2} \left( - \frac{dt^2}{r^{2(d-1)(z-1)/(d-\theta-1)}} + r^{2\theta/(d-\theta-1)} dr^2 + d\vec{x}^2 \right) \ ,
\end{eqnarray}
where, $z$ is the dynamical critical exponent and $\theta$ is known as the hyperscaling violation exponent of the $d$-dimensional boundary theory.\footnote{We have set curvature of space $R=1$ for simplicity.} This background which can be obtained from an Einstein-Maxwell-Dilaton system was proposed in \cite{Huijse:2011ef} as a gravity toy model of a large class of condensed matter systems. Finite temperature behavior of these systems can be studied by introducing a black hole in the bulk. There is a simpler way to write down a hyperscaling violating background with a black hole inside it \cite{Alishahiha:2012qu}
\begin{eqnarray}
&& ds^2 = r^{2\theta/(d-1)} \left(- f(r) \frac{dt^2}{r^{2z}} + \frac{dr^2}{r^2 f(r)} + \frac{d\vec{x}^2}{r^2} \right) \ , \nonumber\\
&& f(r) = 1 - \left(\frac{r}{r_H}\right)^{\gamma} \ ,
\end{eqnarray}
where $\gamma$ is a real constant that we will keep unspecified, $r_H$ is the location of the horizon; the boundary here is located at $r\to 0$. For $\theta=0$, the above metric reduces to finite temperature Lifshitz background \cite{Kachru:2008yh} and $\theta=0, z=1$ is our good old AdS-Schwarzschild. Temperature of the $d$-dimensional boundary theory is given by
\begin{equation}
T= \frac{\gamma}{4 \pi r_H^z}.  
\end{equation} 

In this section, we will show how the techniques developed in the previous sections can easily be generalized by sketching the calculation of entanglement entropy of an infinite rectangular strip (see figure \ref{holo} for a schematic diagram) specified by 
\begin{equation}
x\equiv x^1 \in \left[-\frac{l}{2},\frac{l}{2}\right],~  x^i\in \left[-\frac{L}{2},\frac{L}{2}\right], i=2,...,d-1
\end{equation}
with $L \rightarrow \infty$. We will consider the case $d-\theta-2 \ge 0$; our goal is to demonstrate that the physics is exactly the same as the relativistic case. We will assume that the Rye-Takayanagi prescription for the entanglement entropy holds for these hyperscaling violating backgrounds.

Here we want to note that the results obtained in this section can be used to understand the properties of mutual information which contains rich physics.\footnote{We will investigate this in detail in our future paper \cite{Fischler:2012uv}.}

\subsection{Entanglement entropy: $\theta \not = d-2$}
In this case the divergent part of the entanglement entropy follows an area law \cite{Dong:2012se}:
\begin{equation}
S_{div}=\frac{c}{d-\theta-2} \left(\frac{L^{d-2}}{\epsilon^{d-\theta-2}}\right),
\end{equation}
where
\begin{equation}
c=\frac{1}{4 G_N^{(d+1)}}.
\end{equation}
and $\epsilon$ is the short distance cut-off. Similar to the previous cases, we can formally write down infinite series expansions for $l$ and $S_{A;finite}$ in terms of the closest approach parameter $r_c$
\begin{align}
l=&r_c \sum_{n=0}^{n=\infty}p_n \left(\frac{r_c}{r_H}\right)^{n\gamma},\\
S_{A; \rm finite}=&\frac{2 c\, L^{d-2}}{r_c^{d-\theta-2}} \left[q_0+ \sum_{n=1}^{n=\infty}q_n \left(\frac{r_c}{r_H}\right)^{n\gamma}\right] \ ,
\end{align}
where, $p_n,q_n$ are constants that depend only on $d$ and $\theta$.

\subsubsection*{Zero temperature}
At zero temperature, the entanglement entropy is given by,
\begin{equation}
S_A=S_{div} + \frac{c\, \C(\theta,d) L^{d-2}}{l^{d-\theta-2}}
\end{equation}
where,
\begin{align}
\C(\theta,d)=2 p_0^{d-\theta-2} q_0 \ .
\end{align}

\subsubsection*{Low temperature limit}
At low temperature, similar to the relativistic case, the extremal surface is restricted to be near the boundary region ($r_c\ll r_H$) and hence the leading contribution to the entanglement entropy comes from the boundary. Subleading contributions are small and can be calculated perturbatively 
\begin{equation}
 S_A=S_{div}+\frac{c\, \C(\theta,d) L^{d-2}}{l^{d-\theta-2}}\left[1+h_1~ l^{\gamma}~ T^{\frac{\gamma}{z}}+...\right] \ , 
\end{equation}
where, $h_1$ is a numerical constant.

\subsubsection*{High temperature limit}
At high temperature, $r_c\sim r_H$ and the extremal surface tends to wrap a part of the horizon. Our previous calculations suggest that in the limit $r_c\rightarrow r_H$ we can write
\begin{equation}
 S_{A; \rm finite}=\frac{2 c\, L^{d-2}}{r_H^{d-\theta-2}} \left[q_0-p_0+\frac{l}{r_H}+ \sum_{n=1}^{n=\infty}(q_n-p_n) \right]
\end{equation}
and the infinite sum now converges. Finally, we obtain
\begin{equation} \label{eehighhs}
 S_{A}=S_{div}+ c  \,  L^{d-2}~ T^{\frac{d-\theta-1}{z}}\left[h_2 l  +h_3 ~ T^{-\frac{1}{z}}+...\right] \ , 
\end{equation}
where, $h_2, h_3$ are numerical constants. Similar to the relativistic case, the leading finite part comes from the near horizon part of the geometry and it corresponds to the thermal entropy of region $A$. Full bulk geometry contributes to the finite subleading term and it measures actual quantum entanglement between region $A$ and the surroundings. \\

\subsection{Entanglement entropy: $\theta = d-2$}
\subsubsection*{Zero temperature}
In this case there is a violation of the area law for the divergent part of the entanglement entropy and at zero temperature, the entanglement entropy is given by \cite{Dong:2012se},
\begin{equation}
S_{A}=2 c L^{d-2}\ln\left(\frac{l}{\epsilon}\right).
\end{equation}
The physics at finite temperature is exactly the same as the previous case.
\subsubsection*{Low temperature limit}
At low temperature, $r_c\ll r_H$ and we obtain
\begin{equation}
 S_{A}=2cL^{d-2}\left[ \ln(l/\epsilon)+k_1 l^{\gamma} ~T^{\gamma/z}+...\right] \ ,
\end{equation}
where $k_1\ge 0$ is a numerical constant. 
\subsubsection*{High temperature limit}
At high temperature, $r_c\sim r_H$ and we get
\begin{equation}
 S_{A}=c \,  L^{d-2}\left[- 2\ln(T^{1/z}\epsilon)+k_3 l T^{1/z}+k_2+...\right] \ ,
\end{equation}
where, $k_2$ and $k_3$ are numerical constants.

%%%%%%%%%%%%%%%%%%%%%%%%%%%%%%%%%%%%%%%%%%%%%%%%%%%%%%%%%%%%%%%%%%%%%%%%%%%%%%%%%%%%%%
\section{$\N=4$ super-Yang-Mills in $(3+1)$ dimensions at finite temperature}\label{secex}
In this section, we will use all the tools we have developed in the previous sections to study the prototype case of $\N=4$ super-Yang-Mills with gauge group $U(N)$ at finite temperature. The AdS/CFT correspondence relates this theory to type IIB string theory on asymptotically AdS$_5\times S^5$ spacetime. And as a consequence, in the limit $N\gg1, \lambda=g_{YM}^2 N\gg1$, the theory can be well approximated by the classical supergravity. 

$\N=4$ super-Yang-Mills theory is the unique maximally supersymmetric gauge theory in (3+1) dimensions and it consists of gauge field $A_{\mu}$, six scalar fields $\phi_i, i=1,...,6$ and four Weyl fermions $\chi_\alpha$, all in the adjoint representation of $U(N)$. The Euclidean action for the theory has the schematic form
\begin{align}\label{n4action}
S=&-\frac{1}{4 g^2}\int d^4 x ~\tr \left(F^2 + 2 D_{\mu}\phi_iD^{\mu}\phi_i   +\chi ~\dcut ~\chi + \chi \left[\phi , \chi \right]-\sum_{i,j}[\phi_i,\phi_j]^2\right)\nonumber\\
&+\frac{\theta}{8 \pi^2}\int d^4 x ~\tr\left(F \bar{F}\right).
\end{align}
Now at finite temperature, in the Euclidean description, the system lives on $\mathds{R}^3\times S^1$, the circle direction corresponds to the Euclidean time with period $T^{-1}$. At length scale $l\gg 1/T$, at least for weak coupling, one can perform a Kaluza-Kline reduction along the circle. All the fermions of the theory get a mass of order $T$ at tree level because of the antiperiodic boundary conditions around the circle. The scalars get a mass of the order $g^2 T$ at one loop level through their couplings to fermions. Therefore, at least for weak coupling, at large distance $l\gg 1/T$, the theory reduces to non-supersymmetric Yang-Mills theory in three dimensions. Because the theory has a mass gap, at large temperature ($Tl\gg 1$), it is expected that correlation function $\langle\O(x,t)\O(y,t)\rangle\sim e^{-|x-y|T}$.  

Things are more complicated at strong coupling.  Fortunately, thanks to AdS/CFT, at strong coupling, we can use the supergravity solution to study the system. The mass of the excitations of the QCD strings \cite{Brandhuber:1998er} is $M_s= \lambda^{1/4} T$ and the mass associated with the compactification is $M_c=T$. In the strong coupling region ($\lambda \gg 1$), $M_s\gg M_c$ and as a result QCD strings can ``see" the compactified circle. Therefore at high temperature ($Tl\gg 1$), the theory does not reduce to pure Yang-Mills in three dimensions but $\N=4$ super-Yang-Mills in four dimensions compactified on a circle. 
 
Let us now put $\hbar$ back in the picture for clarity. The action (\ref{n4action}) has a dimension $ML$. The equal time two-point function of the gauge invariant variable $\O=\frac{1}{4}Tr F^2$ is a useful non-local observable to understand the low and high temperature behavior of the theory. The operator $\O$ has conformal dimension $\Delta=4$ with a unit $ML^{-3}$.\footnote{The saddle point approximation that we have used to obtain the two-point function is a good approximation only for large $\Delta$. For definiteness we will use $\O=\frac{1}{4}Tr F^2$ just as an example and consider only the geodesic contributions and ignore the sub-leading corrections.} Now using equation (\ref{ads4low}) and the relation
\begin{equation}
T=\frac{r_H \hbar}{\pi R^2}
\end{equation}
at low temperature ($T|x-y|\ll \hbar$) we obtain
\begin{equation}
\langle\O(x,t)\O(y,t)\rangle=\frac{\hbar^2 \lambda^2}{|x-y|^8}\left[1-\frac{8}{15}\left(\frac{\pi T |x-y|}{2 \hbar}\right)^4+ \frac{48}{175}\left(\frac{\pi T |x-y|}{2 \hbar}\right)^8 +...\right].
\end{equation}
The factor of $\hbar^2$ is there because $\langle\O(x,t)\O(y,t)\rangle$ has  dimension $M^2L^{-6}$. The factor of $\lambda^2$ comes from comparing the result with standard zero temperature result. Similarly, at high temperature ($T|x-y|\gg \hbar$), using equation (\ref{ht}), we obtain
\begin{equation}
\langle\O(x,t)\O(y,t)\rangle = \frac{\A_0 \lambda^2 \pi^8 T^8}{\hbar^6}e^{-4 \pi T |x-y|/\hbar}\left[1+2 \sqrt{2} \E_4 ~e^{-\sqrt{2}\pi T |x-y|/\hbar}+...\right].
\end{equation}
Where $\A_0=4.51$ and $\E_4=2.75$. In the last equation, again we have used dimensional analysis to recover $\hbar$. Indeed, at high temperature, the two-point function decays exponentially.

\begin{figure}[!]
\centering
\includegraphics[width=0.8\textwidth]{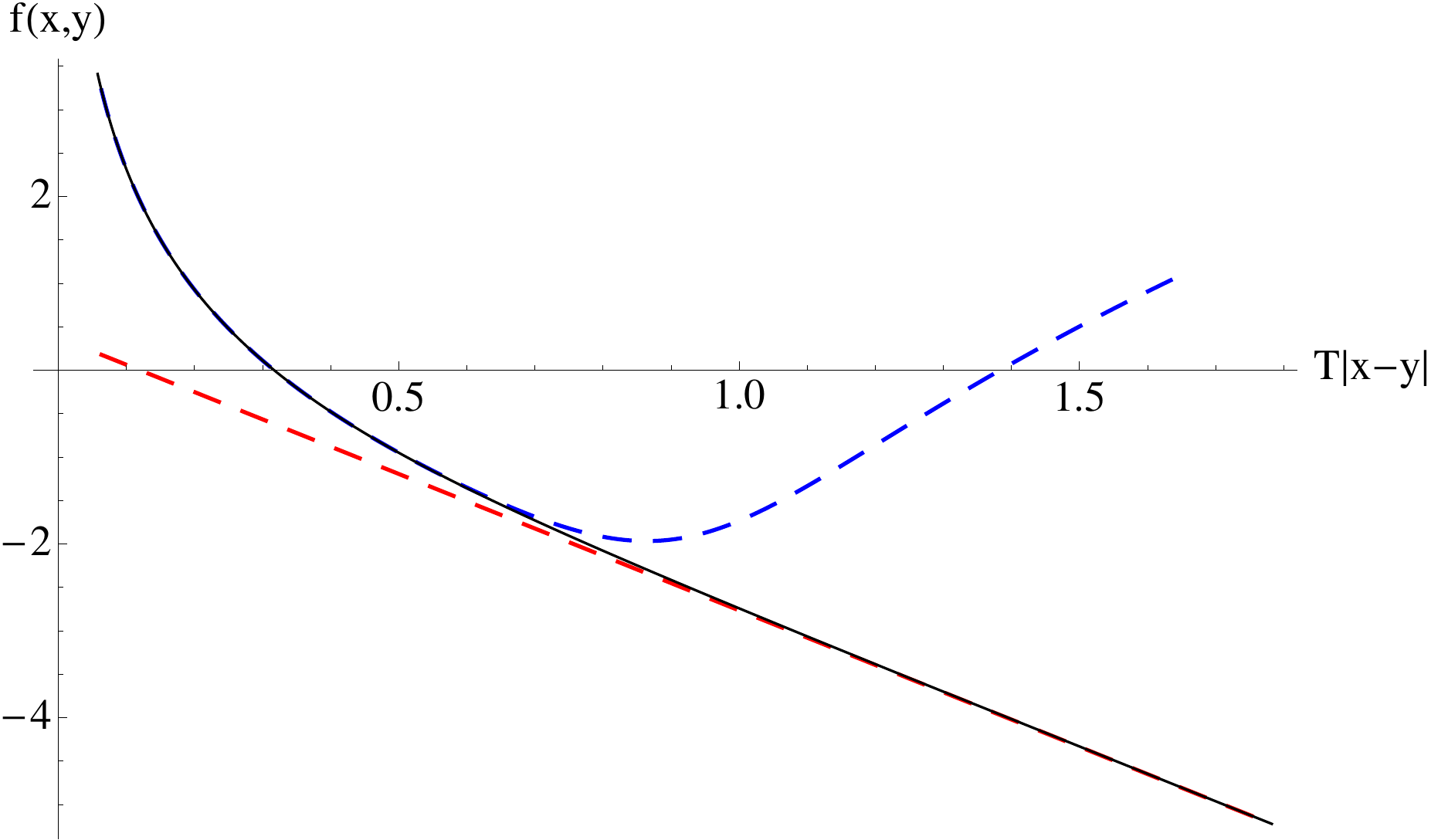}
\caption{Variation of $f(x,y)\equiv \frac{1}{\Delta}\ln \langle \O_{\Delta}(t,x)\O_{\Delta}(t,y)\rangle$ with $T|x-y|$ for the 4-dimensional $\N=4$ SYM theory. The solid black line represents the exact numerical result. Blue and red lines represent the two-point functions computed using low and high temperature approximations, respectively.}
\label{limits}
\end{figure}

Expectation values of the Wilson loops are another set of gauge-invariant non-local observables that are useful to understand the non-perturbative behavior of non-Abelian gauge theories. In this $\N=4$ SYM theory, one should generalize the Wilson loop expression (\ref{wilsonloop}) in the following way
\begin{equation}
W(\C)=\frac{1}{N}Tr\left(\P e^{\oint_{\C}ds\left(A_{\mu}\dot{x}^{\mu}+\vec{n}.\vec{\phi}\sqrt{\dot{x}^2}\right)}\right),
\end{equation}
where, $x^{\mu}(s)$ parametrizes the path $\C$ and $\vec{n}$ is a unit vector in the $\{\phi_i\}$ space. Expectation value of a rectangular infinite Wilson loop with long side of the loop extends along the time direction gives potential energy between a static quark-antiquark pair. Unlike the spatial Wilson loop case a sharp transition takes place at $T\sim 1/l$ for this case \cite{Brandhuber:1998bs}. At low temperature, the U-shaped solutions exist but at high temperature no nontrivial solutions exist. At high temperature, the extremal surface consists of two disjoint vertical surfaces ending at the horizon. Therefore, at high temperature the Wilson loop is independent of separation $l$. In contrast, the transition is more gradual in the case of spatial rectangular (infinite) Wilson loops. In $\N=4$ SYM theory, the expectation value of a spatial rectangular (infinite) Wilson loop is given by,
\begin{equation}
\langle W(\C)\rangle=e^{- S_{NG;ren}}
\end{equation}
and using equation (\ref{lowtempwl}) at low temperature ($Tl\ll \hbar $), we obtain
\begin{align}
S_{NG;ren}=-\frac{4 \pi^2 L  \sqrt{\lambda}}{l \Gamma(\frac{1}{4})^4}\left[1 -\frac{\Gamma \left(\frac{1}{4}\right)^8}{320 \pi ^2} \left(\frac{T l}{\hbar}\right)^4 +\O\left(\frac{T l}{\hbar}\right)^{8}\right].
\end{align} 
At high temperature ($Tl\gg \hbar $), equation (\ref{highwl}) leads to
\begin{equation}
S_{NG;ren}=\frac{\pi A T^2 \sqrt{\lambda}}{2\hbar^2}\left[1-\frac{2 \hbar }{\pi T l}\left(1+\frac{\E_{wl} }{2}~e^{-2\pi l T/\hbar}\right)\right]
\end{equation}
where, $A=lL$ is the area of the loop and $\E_{wl}=1.66$. We have used the fact that for this theory $R^2/ \alpha'=\sqrt{\lambda}$.

\begin{figure}[!]
\centering
\includegraphics[width=0.7\textwidth]{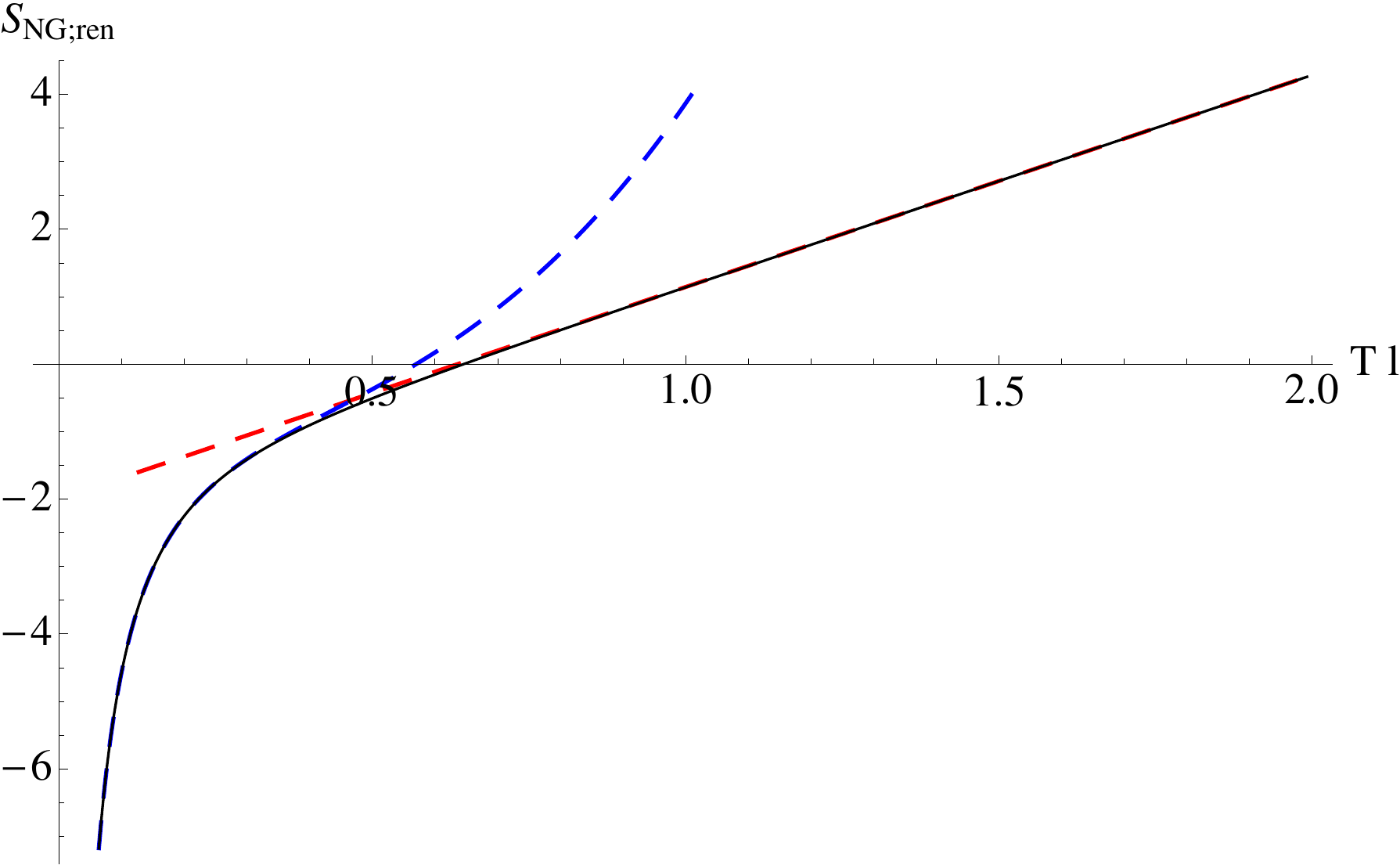}
\caption{Variation of $S_{NG;ren}$ (in the units of $\frac{\sqrt{\lambda} L}{2\pi}$) of the spatial rectangular (infinite) Wilson loop with $T l$ for the 4-dimensional $\N=4$ SYM theory. The expectation value of the Wilson loop is given by $\langle W(\C)\rangle=e^{- S_{NG;ren}}$. The solid black line represents the exact numerical result. Blue and red lines represent the two-point functions computed using low and high temperature approximations, respectively.}
\label{wllimits}
\end{figure}

In a field theory the entanglement entropy is always divergent because there are too many degrees of freedom and we can write
\begin{equation}
S_A=S_{div}+S_{finite}.
\end{equation}
We will consider the entanglement entropy of an infinite rectangular strip (see section \ref{secee}). For the case in hand the divergent piece is temperature independent 
\begin{equation}
S_{div}=\frac{1}{2\pi}N^2 \left(\frac{L}{a}\right)^2.
\end{equation}
Where we have used the AdS/CFT dictionary to write $R^3/4G_N^{(1+4)}=N^2/2\pi$. At low temperature ($Tl\ll \hbar $), the finite part is given by (see equation (\ref{eel})) 
\begin{equation}
S_{finite}=-N^2 (0.051) \left(\frac{L}{l}\right)^{2}\left[1-(1.764) \left(\frac{\pi T l}{\hbar}\right)^4+ \O\left(\frac{\pi T l}{\hbar}\right)^{8}\right].
\end{equation}
And at high temperature ($Tl\gg \hbar $), following equation (\ref{eeh}) we obtain
\begin{equation}
S_{finite}=\frac{\pi^2 N^2 }{2 \hbar^3}\left[VT^3- (0.106)\hbar A T^2 -\frac{\E_{ent}}{2\pi} \sqrt{\frac{3}{2}}\hbar A T^2 e^{-\sqrt{6} \pi T l/\hbar}+...\right]
\end{equation}
where, $\E_{ent}=1.174$, $V=l L^2$ is the volume of the rectangular strip and $A=2 L^2$ is the area. Note that the leading behavior is consistent with the observation of \cite{Ryu:2006ef}.

\begin{figure}[!]
\centering
\includegraphics[width=0.8\textwidth]{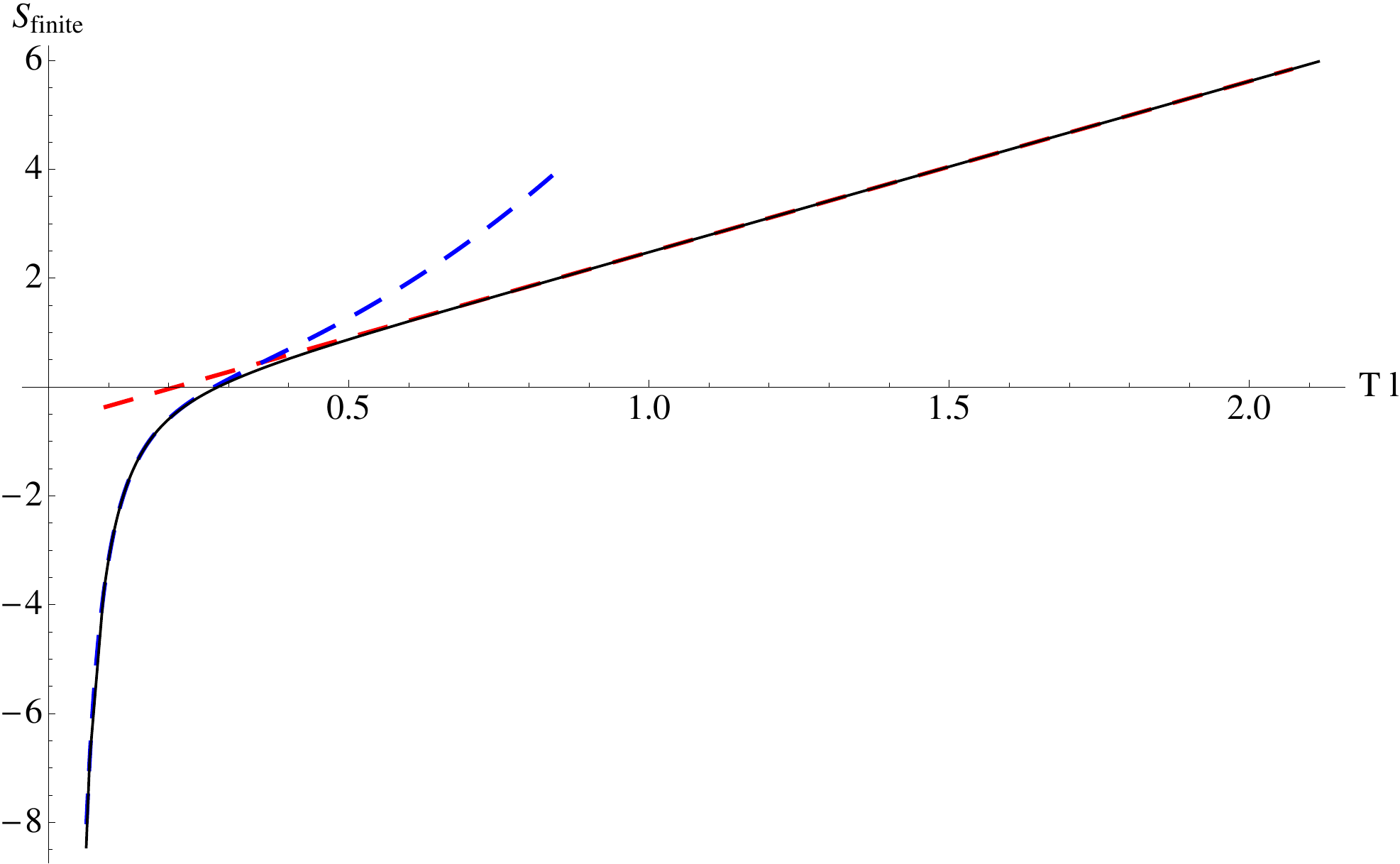}
\caption{Variation of $S_{finite}$, the finite part of the entanglement entropy (in the units of $\frac{N^2 L^2}{2\pi}$) of an infinite rectangular strip with $T l$ for the 4-dimensional $\N=4$ SYM theory.  The solid black line represents the exact numerical result. Blue and red lines represent the two-point functions computed using low and high temperature approximations, respectively.}
\label{eelimits}
\end{figure}
%%%%%%%%%%%%%%%%%%%%%%%%%%%%%%%%%%%%%%%%%%%%%%%%%%%%%%%%%%%%%%%%%%%%%%%%%%%%%%%%%%%%
\section{Conclusions}\label{conclusions}
In this article, we have studied the high and low temperature behavior of some non-local observables in strongly coupled large-N gauge theories. At low temperature, the leading term is obviously the zero temperature contribution and the sub-leading terms indicate a decrease in correlations as the system is heated. At high temperature, the leading contributions come from the near horizon part of the bulk. The full bulk geometry contributes significantly to the sub-leading terms. We have investigated the contributions from different regions of the bulk  for the equal time two-point function, rectangular (infinite) spatial Wilson loop and entanglement entropy of an infinite rectangular strip. In the case of spatial Wilson loops and the entanglement entropy, our calculations suggest that the general form of the high temperature answer does not particularly depend on the shape. In particular, it is interesting that at high temperature we observe an area law behavior of the finite subleading term of the entanglement entropy for both relativistic and non-relativistic theories.  This is consistent with what is expected from Quantum Information Theory when the two-point function decays \cite{Wolf:2008}; however, the Quantum Information Theory argument is qualitative and can not be extended easily to non-zero temperature. On the other hand, mutual information \cite{Wolf:2008} is another important concept in information theory which is well understood. Techniques developed in this paper can be used to study the behavior of mutual information at nonzero temperature \cite{Fischler:2012uv}  for a large class of field theories with holographic duals. 

In forthcoming work we are generalizing this approach to include finite densities. We also plan to extend our techniques to the analysis of holographic thermalization which has attracted much attention in recent years \cite{AbajoArrastia:2010yt, Aparicio:2011zy, Albash:2010mv, Balasubramanian:2010ce, Balasubramanian:2011ur, Galante:2012pv, Caceres:2012em}. An analytic understanding of thermalization at different limits can possibly provide new insight into the non-equilibrium physics.

\section*{Acknowledgements}
We are grateful to Matt Headrick and Arnab Kundu for helpful discussions. We would like to thank  Juan F. Pedraza for pointing out an error. This material is based upon work supported by the National Science Foundation under Grant Number PHY-0969020 and by the Texas Cosmology Center, which is supported by the College of Natural Sciences and the Department of Astronomy at the University of Texas at Austin and the McDonald Observatory.

%%%%%%%%%%%%%%%%%%%%%%%%%%%%%%%%%%%%%%%%%%%%%%%%%%%%%%%%%%%%%%%%%%%%%%%%%%%%%%%%%%%%%%%%%
\appendix

\section{Two-point function: Low temperature expansion}\label{lte}
At low temperature, $r_H/r_c\ll 1$. From equation (\ref{seriesrc}), keeping only few subleading terms, we get

\begin{align}
l=\frac{1}{r_c}\left[2+ \frac{\sqrt{\pi } \Gamma \left(\frac{d}{2}+1\right)}{2 \Gamma \left(\frac{d+3}{2}\right)}\left(\frac{r_H}{r_c}\right)^d+\frac{3 \sqrt{\pi } \Gamma (d+1)}{8 \Gamma \left(d+\frac{3}{2}\right)}\left(\frac{r_H}{r_c}\right)^{2d}+...\right].
\end{align} 
Solving this perturbatively we obtain
\begin{align}
r_c=\frac{1}{l}&\left[2+ \frac{\sqrt{\pi } \Gamma \left(\frac{d}{2}+1\right)}{2 \Gamma \left(\frac{d+3}{2}\right)}\left(\frac{r_H l}{2}\right)^d\right.\nonumber\\
&\left.+\left(\frac{3 \sqrt{\pi } \Gamma (d+1)}{8 \Gamma \left(d+\frac{3}{2}\right)}-\frac{\pi d}{8}\left(\frac{ \Gamma \left(\frac{d}{2}+1\right)}{ \Gamma \left(\frac{d+3}{2}\right)}\right)^2 \right)\left(\frac{r_H l}{2}\right)^{2d}+\O\left(\frac{r_H l}{2}\right)^{3d}\right].\label{lowtemprc}
\end{align}
In this low temperature limit, equation (\ref{sreneqn}) becomes
\begin{align}
\L_{ren}=& 2\ln\left(\frac{2}{r_c}\right)+\frac{\sqrt{\pi } \Gamma \left(\frac{d}{2}\right)}{2 \Gamma \left(\frac{d+1}{2}\right)}\left(\frac{r_H}{r_c}\right)^{d}+\frac{3 \sqrt{\pi } \Gamma (d)}{8 \Gamma \left(d+\frac{1}{2}\right)}\left(\frac{r_H}{r_c}\right)^{2d}+ \O\left(\frac{r_H}{r_c}\right)^{3d}.
\end{align}
Now using equation (\ref{lowtemprc}), we get 
\begin{align}
\L_{ren}= 2\ln\left(l\right)+\frac{\sqrt{\pi } \Gamma \left(\frac{d}{2}\right)}{4 \Gamma \left(\frac{d+3}{2}\right)}\left(\frac{r_H l}{2}\right)^d
+& \left(\frac{3 \sqrt{\pi } \Gamma (d)}{16 \Gamma \left(d+\frac{3}{2}\right)}-\frac{\pi  \Gamma \left(\frac{d}{2}+1\right)^2}{16 \Gamma \left(\frac{d+3}{2}\right)^2}\right) \left(\frac{r_H l}{2}\right)^{2d}\nonumber\\
+&\O\left(\frac{r_H l}{2}\right)^{3d}.
\end{align}
Therefore, the two-point function is
\begin{align}
\langle \O(t,x)\O(t,y)\rangle=\frac{1}{l^{2 \Delta}}\left[1+\C_1 \left(\frac{r_H l}{2}\right)^d + \C_2 \left(\frac{r_H l}{2}\right)^{2d}+\O\left(\frac{r_H l}{2}\right)^{3d}\right]\label{lowtemp2}
\end{align}
where again $l=|x-y|$ and 
\begin{align}
\C_1=&-\frac{ \sqrt{\pi } \Delta  \Gamma \left(\frac{d}{2}\right)}{4 \Gamma \left(\frac{d+3}{2}\right)},\label{c1}\\
\C_2=&\frac{1}{64} \left[\pi  \Delta  \left(d^2+2 \Delta \right)\left(\frac{ \Gamma \left(\frac{d}{2}\right)}{\Gamma \left(\frac{d+3}{2}\right)}\right)^2-\frac{12 \sqrt{\pi } \Delta  \Gamma (d)}{\Gamma \left(d+\frac{3}{2}\right)}\right].\label{c2}
\end{align}
%%%%%%%%%%%%%%%%%%%%%%%%%%%%%%%%%%%%%%%%%%%%%%%%%%%%%%%%%%%%%%%%%%%%%%%%%%%%%%%%%%%%%%%
\section{High temperature subleading term}\label{subleading}
\subsection{Two-point function}
We have mentioned that at high temperature $r_c\sim r_H$. Now we will try to figure out how fast $r_c$ approaches $r_H$. For large $n$, the series (\ref{seriesrc}) goes as $\sim \frac{1}{n}\left(r_H/r_c\right)^{nd}$ and hence it diverges for $r_c=r_H$. We can isolate the divergent piece from the infinite series (\ref{seriesrc}) , yielding
\begin{align}
l=\frac{1}{ r_c}\sum_{n=1}^\infty\left(\frac{\Gamma\left[\frac{1}{2}+n\right]\Gamma\left[1+\frac{nd}{2}\right]}{ \Gamma[1+n]\Gamma\left[\frac{3}{2}+\frac{nd}{2}\right]}-\frac{\sqrt{2}}{\sqrt{d}n}\right)\left(\frac{r_H}{r_c}\right)^{nd}+\frac{2}{r_c} -\frac{\sqrt{2}}{\sqrt{d}r_c}\ln\left[1-\left(\frac{r_H}{r_c}\right)^{d}\right].
\end{align}
The infinite series in the last equation now converges even for $r_c=r_H$. It is convenient to write $r_c=r_H(1+\epsilon)$ for the purpose of solving the last equation. It is very clear from previous discussions that at high temperature $\epsilon\ll 1$. Therefore, from the last expression, we obtain
\begin{align}
\frac{\sqrt{2}}{\sqrt{d}}\ln\left[\epsilon d\right]=-lr_H+2+ \sum_{n=1}^\infty\left(\frac{\Gamma\left[\frac{1}{2}+n\right]\Gamma\left[1+\frac{nd}{2}\right]}{ \Gamma[1+n]\Gamma\left[\frac{3}{2}+\frac{nd}{2}\right]}-\frac{\sqrt{2}}{\sqrt{d}n}\right)+\O(\epsilon).
\end{align}
Solving the last equation for $\epsilon$, at the leading order we get
\begin{equation}\label{epsilon}
\epsilon=\E_d ~e^{-\sqrt{\frac{d}{2}}l r_H},
\end{equation}
where $\E_d$ is given by,
\begin{equation}\label{ed}
\E_d=\frac{1}{d}\exp\left[\sqrt{\frac{d}{2}}\left\{2+\sum_{n=1}^\infty\left(\frac{\Gamma\left[\frac{1}{2}+n\right]\Gamma\left[1+\frac{nd}{2}\right]}{ \Gamma[1+n]\Gamma\left[\frac{3}{2}+\frac{nd}{2}\right]}-\frac{\sqrt{2}}{\sqrt{d}n}\right)\right\}\right].
\end{equation}

Now we will compute the next order correction to the high temperature result (\ref{hightemp}). First we will start with equation (\ref{modsren})
\begin{align}
\L_{ren}=&  2\ln\left(\frac{2}{r_c}\right)+\left(r_c l -2\right)+\sum_{n=1}^\infty\left(\frac{1}{nd}\right)\frac{\Gamma\left[\frac{1}{2}+n\right]\Gamma\left[1+\frac{nd}{2}\right]}{ \Gamma[1+n]\Gamma\left[\frac{3}{2}+\frac{nd}{2}\right]}\left(\frac{r_H}{r_c}\right)^{nd}.\nonumber
\end{align}
We can not yet write $r_c=r_H(1+\epsilon)$ and do an expansion for small $\epsilon$. The term linear in $\epsilon$ contains an infinite series that does not converge. Therefore, in order to obtain a finite first order expression in $\epsilon$, we will write the last equation in the following way
\begin{align}
\L_{ren}=  2\ln\left(\frac{2}{r_c}\right)+\left(r_c l -2\right)&+\sum_{n=1}^\infty\left[\left(\frac{1}{nd}\right)\frac{\Gamma\left[\frac{1}{2}+n\right]\Gamma\left[1+\frac{nd}{2}\right]}{ \Gamma[1+n]\Gamma\left[\frac{3}{2}+\frac{nd}{2}\right]}-\frac{\sqrt{2}}{d^{3/2}n^2}\right]\left(\frac{r_H}{r_c}\right)^{nd}\nonumber\\
&+\frac{\sqrt{2}}{d^{3/2}}\text{Li}_2\left[\left(\frac{r_H}{r_c}\right)^{d}\right],
\end{align}
where, Li is the PolyLog function. We have summed the part of the infinite series that could diverge in the first oder in $\epsilon$. Now, we can write $r_c=r_H(1+\epsilon)$ and do an expansion. In the first order in $\epsilon$ we obtain,
\begin{align}
\L_{ren}=  &2\ln\left(\frac{2}{r_H}\right)+\left(r_H l -2\right)+\sum_{n=1}^\infty\left(\frac{1}{nd}\right)\frac{\Gamma\left[\frac{1}{2}+n\right]\Gamma\left[1+\frac{nd}{2}\right]}{ \Gamma[1+n]\Gamma\left[\frac{3}{2}+\frac{nd}{2}\right]}\nonumber\\
&-2\epsilon+r_H l \epsilon-\epsilon \sum_{n=1}^\infty\left[\frac{\Gamma\left[\frac{1}{2}+n\right]\Gamma\left[1+\frac{nd}{2}\right]}{ \Gamma[1+n]\Gamma\left[\frac{3}{2}+\frac{nd}{2}\right]}-\frac{\sqrt{2}}{\sqrt{d}n}\right]+\frac{\sqrt{2}}{\sqrt{d}}\left(-1+\log(\epsilon d)\right)\epsilon\nonumber\\
&+\O(\epsilon^2).
\end{align}
In the first line, we have all the terms that contribute in the leading order at high temperature (we will call it $\L_{leading}$) and in the second line we have all the subleading terms. We can simplify the second line farther by using equation (\ref{epsilon})
\begin{equation}
\L_{ren}=\L_{leading}-\sqrt{\frac{2}{d}}\epsilon +\O(\epsilon^2).
\end{equation}
Therefore, using (\ref{twopoint}) we finally obtain
\begin{align}\label{hightemp1}
\langle \O(t,x)\O(t,y)\rangle = \A_{d,\Delta}~ r_H^{2 \Delta}~ e^{-\Delta r_H l}\left[1+\sqrt{\frac{2}{d}}\Delta \E_d ~e^{-\sqrt{\frac{d}{2}}l r_H}+...\right],
\end{align}
where $\E_d$ is given by equation (\ref{ed}) and the dots represent the higher order correction terms.

\subsection{Rectangular Wilson loop}
Again we will isolate the divergent part of the series (\ref{wlrc})
\begin{align}
l=\frac{1}{2 r_c}\sum_{n=1}^\infty &\left(\frac{\Gamma\left[\frac{1}{2}+n\right]\Gamma\left[\frac{1}{4}(3+nd)\right]}{ \Gamma[1+n]\Gamma\left[\frac{1}{4}(5+nd)\right]}-\frac{2}{\sqrt{d} n}\right)\left(\frac{r_H}{r_c}\right)^{nd}\nonumber\\
&+\frac{2\sqrt{2}\pi^{3/2}}{r_c \Gamma(\frac{1}{4})^2}-\frac{1}{\sqrt{d}r_c}\ln\left[1-\left(\frac{r_H}{r_c}\right)^{d}\right].
\end{align}
It is again convenient to write $r_c=r_H(1+\epsilon)$ and at high temperature $\epsilon\ll 1$. Therefore, from the last expression, we obtain
\begin{align}
\frac{1}{\sqrt{d}}\ln\left[\epsilon d\right]=-lr_H+\frac{2\sqrt{2}\pi^{3/2}}{ \Gamma(\frac{1}{4})^2}+ \frac{1}{2 }\sum_{n=1}^\infty \left(\frac{\Gamma\left[\frac{1}{2}+n\right]\Gamma\left[\frac{1}{4}(3+nd)\right]}{ \Gamma[1+n]\Gamma\left[\frac{1}{4}(5+nd)\right]}-\frac{2}{\sqrt{d} n}\right)+\O(\epsilon).
\end{align}
Solving the last equation for $\epsilon$, at the leading order we get
\begin{equation}\label{epsilonwl}
\epsilon=\E_{wl} ~e^{-\sqrt{d}~l r_H},
\end{equation}
where $\E_{wl}$ is a constant given by,
\begin{equation}
\E_{wl}=\frac{1}{d}\exp\left[\sqrt{d}\left\{\frac{2\sqrt{2}\pi^{3/2}}{ \Gamma(\frac{1}{4})^2}+ \frac{1}{2 }\sum_{n=1}^\infty \left(\frac{\Gamma\left[\frac{1}{2}+n\right]\Gamma\left[\frac{1}{4}(3+nd)\right]}{ \Gamma[1+n]\Gamma\left[\frac{1}{4}(5+nd)\right]}-\frac{2}{\sqrt{d} n}\right)\right\}\right].
\end{equation}

Now we will compute the next order correction to high temperature result (\ref{hightempwl}). First we will start with
\begin{align}
S_{NG;ren}=\frac{L r_c }{\pi \alpha'}\left[-\frac{2\sqrt{2}\pi^{3/2}}{\Gamma(\frac{1}{4})^2}+\frac{l r_c}{2}+\frac{1}{2}\sum_{n=1}^\infty\frac{1}{nd-1}\frac{\Gamma\left[\frac{1}{2}+n\right]\Gamma\left[\frac{1}{4}(3+nd)\right]}{ \Gamma[1+n]\Gamma\left[\frac{1}{4}(5+nd)\right]}\left(\frac{r_H}{r_c}\right)^{nd}\right].
\end{align}
Before, we write $r_c=r_H(1+\epsilon)$ and do an expansion for small $\epsilon$, in order to obtain a finite first order expression in $\epsilon$, we will write the last equation in the following way
\begin{align}
S_{NG;ren}=&\frac{L r_c }{\pi \alpha'}\left[-\frac{2\sqrt{2}\pi^{3/2}}{\Gamma(\frac{1}{4})^2}+\frac{l r_c}{2}+\frac{1}{d^{3/2}}\text{Li}_2\left\{\left(\frac{r_H}{r_c}\right)^{d}\right\}\right.\nonumber\\
& \left.+\frac{1}{2}\sum_{n=1}^\infty\left(\frac{1}{nd-1}\frac{\Gamma\left[\frac{1}{2}+n\right]\Gamma\left[\frac{1}{4}(3+nd)\right]}{ \Gamma[1+n]\Gamma\left[\frac{1}{4}(5+nd)\right]}-\frac{2}{d^{3/2}n^2}\right)\left(\frac{r_H}{r_c}\right)^{nd}\right].
\end{align}
Where, Li is the PolyLog function. We have again summed the part of the infinite series that could diverge at first order in $\epsilon$. Now, we will write $r_c=r_H(1+\epsilon)$ and do an expansion. At first order in $\epsilon$ we obtain,
\begin{align}
S_{NG;ren}=&\frac{L r_H }{\pi \alpha'}\left[-\frac{2\sqrt{2}\pi^{3/2}}{\Gamma(\frac{1}{4})^2}+\frac{l r_H}{2}+\frac{1}{2}\sum_{n=1}^\infty\frac{1}{nd-1}\frac{\Gamma\left[\frac{1}{2}+n\right]\Gamma\left[\frac{1}{4}(3+nd)\right]}{ \Gamma[1+n]\Gamma\left[\frac{1}{4}(5+nd)\right]}\right]\nonumber\\
&+\frac{\epsilon L r_H }{\pi \alpha'}\left[-\frac{2\sqrt{2}\pi^{3/2}}{\Gamma(\frac{1}{4})^2}+lr_H+\frac{1}{\sqrt{d}}\left(-1+\log(\epsilon d)\right)\right.\nonumber\\
&\left.-\frac{1}{2}\sum_{n=1}^\infty\left(\frac{\Gamma\left[\frac{1}{2}+n\right]\Gamma\left[\frac{1}{4}(3+nd)\right]}{ \Gamma[1+n]\Gamma\left[\frac{1}{4}(5+nd)\right]}-\frac{2}{d^{1/2}n}\right)\right]+\O(\epsilon^2).
\end{align}
In the first line, we have all the terms that contribute in the leading order at high temperature (we will call it $S_{NG;leading}$). We can simplify the subleading terms farther by using equation (\ref{epsilonwl})
\begin{equation}
S_{NG;ren}=S_{NG;leading}-\frac{\epsilon L r_H }{\pi \alpha'\sqrt{d}} +\O(\epsilon^2).
\end{equation}
Therefore, finally we have
\begin{align}
S_{NG;ren}=\frac{L r_H }{\pi \alpha'}\left[\W_d+\frac{l r_H}{2}-\frac{1}{\sqrt{d}}\E_{wl} ~e^{-\sqrt{d}~l r_H}+...\right].
\end{align}
%%%%%%%%%%%%%%%%%%%%%%%%%%%%%%%%%%%%%%%%%%%%%%%%%%%%%%%%%%%%%%%%%%%%%%%
\subsection{Entanglement entropy}
We know that at high temperature $r_c\sim r_H$. Now again we will try to figure out how fast $r_c$ approaches $r_H$. For large $n$, the series (\ref{eerc}) goes as $\sim \frac{1}{n}\left(r_H/r_c\right)^{nd}$ and hence it diverges for $r_c=r_H$. We can isolate the divergent piece from the infinite series (\ref{eerc}), obtaining
\begin{align}
l=&\frac{2}{ r_c}\sum_{n=1}^\infty\left\{\left(\frac{1}{1+nd}\right)\frac{\Gamma\left[\frac{1}{2}+n\right]\Gamma \left[\frac{d (n+1)}{2 (d-1)}\right]}{ \Gamma[1+n]\Gamma \left[\frac{d n+1}{2 (d-1)}\right]}-\frac{1}{\sqrt{2} \sqrt{(d-1) d}~ n}\right\}\left(\frac{r_H}{r_c}\right)^{nd}\nonumber\\
&+\frac{2}{ r_c}\frac{\sqrt{\pi } \Gamma \left(\frac{d}{2 (d-1)}\right)}{\Gamma \left(\frac{1}{2 (d-1)}\right)}-\frac{\sqrt{2}}{r_c \sqrt{(d-1) d}}\ln\left[1-\left(\frac{r_H}{r_c}\right)^{d}\right].
\end{align}
The infinite series in the last equation now converges even for $r_c=r_H$. Analogous to the last two cases, now it is convenient to write $r_c=r_H(1+\epsilon)$. It is very clear from previous discussions that at high temperature $\epsilon\ll 1$. Therefore, from the last expression, we obtain
\begin{align}
\frac{\sqrt{2}}{ \sqrt{(d-1) d}}\ln\left[\epsilon d\right]&=-lr_H+\frac{2\sqrt{\pi } \Gamma \left(\frac{d}{2 (d-1)}\right)}{\Gamma \left(\frac{1}{2 (d-1)}\right)}\nonumber\\
&+ 2\sum_{n=1}^\infty\left\{\left(\frac{1}{1+nd}\right)\frac{\Gamma\left[\frac{1}{2}+n\right]\Gamma \left[\frac{d (n+1)}{2 (d-1)}\right]}{ \Gamma[1+n]\Gamma \left[\frac{d n+1}{2 (d-1)}\right]}-\frac{1}{\sqrt{2} \sqrt{(d-1) d}~ n}\right\}+\O(\epsilon).
\end{align}
Solving the last equation for $\epsilon$, at the leading order we get
\begin{equation}\label{epsilonee}
\epsilon=\E_{ent} ~e^{-\sqrt{d(d-1)/2}~l r_H},
\end{equation}
where $\E_{ent}$ is a constant given by,
\begin{align}
\E_{ent}=&\frac{1}{d}\exp\left[\sqrt{\frac{d(d-1)}{2}}\left\{\frac{2\sqrt{\pi } \Gamma \left(\frac{d}{2 (d-1)}\right)}{\Gamma \left(\frac{1}{2 (d-1)}\right)}\right.\right.\nonumber\\
&\left.\left.+ 2\sum_{n=1}^\infty\left\{\left(\frac{1}{1+nd}\right)\frac{\Gamma\left[\frac{1}{2}+n\right]\Gamma \left[\frac{d (n+1)}{2 (d-1)}\right]}{ \Gamma[1+n]\Gamma \left[\frac{d n+1}{2 (d-1)}\right]}-\frac{1}{\sqrt{2} \sqrt{(d-1) d}~ n}\right\}\right\}\right].
\end{align}

Now we will compute the next order correction to the high temperature result (\ref{Shigh}). First we will start with equation (\ref{highA})
\begin{align}
A_{finite}=&2 L^{d-2}r_c^{d-2} \left[\frac{\sqrt{\pi } \Gamma \left(-\frac{d-2}{2 (d-1)}\right)}{2 (d-1) \Gamma \left(\frac{1}{2(d-1)}\right)}+\frac{l r_c}{2}-\frac{\sqrt{\pi } \Gamma \left(\frac{d}{2 (d-1)}\right)}{\Gamma \left(\frac{1}{2(d-1)}\right)}\right.\nonumber\\
&+\left.\sum_{n=1}^\infty\left(\frac{1}{1+nd}\right)\left(\frac{d-1}{d(n-1)+2}\right)\frac{\Gamma\left[\frac{1}{2}+n\right]\Gamma \left[\frac{d (n+1)}{2 d-2}\right]}{ \Gamma[1+n]\Gamma \left[\frac{d n+1}{2 (d-1)}\right]}\left(\frac{r_H}{r_c}\right)^{nd}\right].
\end{align}
We will write $r_c=r_H(1+\epsilon)$ and do an expansion for small $\epsilon$ and then we will use equation (\ref{epsilonee}). But before that we should note that the term linear in $\epsilon$ contains an infinite series that does not converge. We will isolate and resum part of the last equation that could diverge at first order in $\epsilon$, in the following way
\begin{align}
A_{finite}&=2 L^{d-2}r_c^{d-2} \left[\frac{\sqrt{\pi } \Gamma \left(-\frac{d-2}{2 (d-1)}\right)}{2 (d-1) \Gamma \left(\frac{1}{2(d-1)}\right)}+\frac{l r_c}{2}-\frac{\sqrt{\pi } \Gamma \left(\frac{d}{2 (d-1)}\right)}{\Gamma \left(\frac{1}{2(d-1)}\right)}+\sqrt{\frac{d-1}{2 d^3}}\text{Li}_2\left[\left(\frac{r_H}{r_c}\right)^{d}\right]\right.\nonumber\\
&+\left.\sum_{n=1}^\infty\left\{\left(\frac{1}{1+nd}\right)\left(\frac{d-1}{d(n-1)+2}\right)\frac{\Gamma\left[\frac{1}{2}+n\right]\Gamma \left[\frac{d (n+1)}{2 d-2}\right]}{ \Gamma[1+n]\Gamma \left[\frac{d n+1}{2 (d-1)}\right]}-\sqrt{\frac{d-1}{2 d^3}}\frac{1}{n^2}\right\}\left(\frac{r_H}{r_c}\right)^{nd}\right].
\end{align}
Where, Li is the PolyLog function. Now, we will write $r_c=r_H(1+\epsilon)$ and do an expansion. In the first order in $\epsilon$ we obtain,
\begin{align}
A_{finite}&= l L^{d-2}r_H^{d-1}\left[1+ \left(\frac{1}{l r_H}\right)\S_{high}\right]\nonumber\\
&+2 L^{d-2}r_H^{d-2}\epsilon\left[-\frac{\sqrt{\pi}(d-1)\Gamma \left(\frac{d}{2 (d-1)}\right)}{\Gamma \left(\frac{1}{2 (d-1)}\right)} +\frac{d-1}{2}lr_H+\sqrt{\frac{d-1}{2d}}\left(-1+\log(\epsilon d)\right)\right.\nonumber\\
&\left.-(d-1)\sum_{n=1}^\infty\left\{\left(\frac{1}{1+nd}\right)\frac{\Gamma\left[\frac{1}{2}+n\right]\Gamma \left[\frac{d (n+1)}{2 (d-1)}\right]}{ \Gamma[1+n]\Gamma \left[\frac{d n+1}{2 (d-1)}\right]}-\frac{1}{\sqrt{2} \sqrt{(d-1) d}~ n}\right\}\right]+\O(\epsilon^2).
\end{align}
In the first line, we have all the terms that contribute in the leading order at high temperature and $\S_{high}$ is given by equation (\ref{constantshigh}). We can simplify the subleading terms farther by using equation (\ref{epsilonee})
\begin{equation}
A_{finite}= l L^{d-2}r_H^{d-1}\left[1+ \left(\frac{1}{l r_H}\right)\S_{high}\right]-L^{d-2}r_H^{d-2} \sqrt{\frac{2(d-1)}{d}} \epsilon+\O(\epsilon^2).
\end{equation}
Therefore, finally we have
\begin{align}
A_{finite}= V r_H^{d-1}\left[1+ \left(\frac{1}{l r_H}\right)\left(\S_{high}-\sqrt{\frac{2(d-1)}{d}} \E_{ent} ~e^{-\sqrt{d(d-1)/2}~l r_H}\right)+...\right].
\end{align}
%%%%%%%%%%%%%%%%%%%%%%%%%%%%%%%%%%%%%%%%%%%%%%%%%%%%%%%%%%%%%%%%%%%%%%%%%%%%%%%%%%%%%%%

%%%%%%%%%%%%%%%%%%%%%%%%%%%%%%%%%%%%%%%%%%%%%%%%%%%%%%%%%%%%%%%%%%%%%%%%%%%%%%%%%%%%%%%

\end{document}